\documentclass{article}
\usepackage{mathrsfs}
\usepackage{mathtools}
\usepackage{commath}
\usepackage{cite} 
\usepackage{url,ulem,comment}

\usepackage{booktabs} 
\usepackage{microtype} 
\usepackage[colorlinks=false]{hyperref}
\usepackage{listings} 
\usepackage{amsmath,bm}
\usepackage{amssymb}
\usepackage{stackrel}
\usepackage{setspace}
\usepackage{caption}
\usepackage{units}
\usepackage{url}
\usepackage{tikz,graphics,color,fullpage,float,epsf,caption,subcaption}
\usepackage{setspace}
\usepackage[document]{ragged2e}
\usepackage{algorithm,algpseudocode}

\usepackage{amsthm}

\usepackage[document]{ragged2e}

\usepackage[round]{natbib}
\usepackage{lineno}

\newcommand{\btheta}{ \mbox{\boldmath $\theta$}}
\newcommand{\bmu}{ \mbox{\boldmath $\mu$}}

\newcommand{\bbeta}{ \mbox{\boldmath $\beta$}}

\newcommand{\bepsilon}{ \mbox{\boldmath $\epsilon$}}

\newcommand{\bX}{ \mbox{\bf X}}

\newcommand{\bY}{ \mbox{\bf Y}}

\newcommand{\bh}{ \mbox{\bf h}}

\newcommand{\bs}{ \mbox{\bf s}}

\newcommand{\bw}{ \mbox{\bf w}}

\newcommand{\beq}{ \begin{equation}}
\newcommand{\eeq}{ \end{equation}}
\newcommand{\beqn}{ \begin{eqnarray}}
\newcommand{\eeqn}{ \end{eqnarray}}

\usepackage{xcolor}

\title{A comparison between geostatistical and machine learning models for spatio-temporal prediction of PM$_{2.5}$ data}

\author{
Zeinab Mohamed \thanks{Department of Mathematics,
Oberlin College and Conservatory}
\and Wenlong Gong  \thanks{Department of Mathematics and Statistics, University of Houston - Downtown} 
}
\begin{document} 
\maketitle
\begin{justify}

\begin{abstract}
Ambient air pollution poses significant health and environmental challenges. Exposure to high concentrations of PM$_{2.5}$ has been linked to increased respiratory and cardiovascular hospital admissions, more emergency department visits and deaths. Traditional air quality monitoring systems such as EPA-certified stations provide limited spatial and temporal data. The advent of low-cost sensors has dramatically improved the granularity of air quality data, enabling real-time, high-resolution monitoring. This study exploits the extensive data from PurpleAir sensors to assess and compare the effectiveness of various statistical and machine learning models in producing accurate hourly PM$_{2.5}$ maps across California. We evaluate traditional geostatistical methods, including universal kriging, nearest neighbor Gaussian process and fixed rank kriging, against advanced machine learning approaches such as neural network, random forest, and support vector machine, as well as ensemble model. Our findings highlight the synergistic value of combining geostatistical methods with machine learning to achieve higher-accuracy PM$_{2.5}$ predictions with lower computational burden, while flexibly capturing nonlinear structure when present.
\end{abstract}


\section{Introduction}

The disparate distribution of air pollution underscores significant ecological justice concerns \citep{hajat2015socioeconomic}, particularly for fine particulate matter with an aerodynamic diameter less than 2.5 micrometers (PM$_{2.5}$). As one of the most hazardous air pollutants, PM${2.5}$ is associated with substantial adverse health effects and contributes to broader environmental risks {\citep{world2018noncommunicable}. In urban settings, \textcolor{black}{the government typically operates a limited number of United States Environmental Protection Agency (EPA)–certified air quality monitoring stations, which report hourly averaged pollutant concentrations.} However, air pollution levels can vary substantially over short temporal scales and within small spatial areas. The emergence of low-cost air pollution sensors, such as \textcolor{black}{PurpleAir (PA) sensors}, has substantially expanded monitoring coverage by providing measurements at high spatial and temporal resolution. These data have improved the identification of localized pollution hotspots and enhanced the assessment of air quality conditions often summarized using the \textcolor{black}{Air Quality Index (AQI)} during extreme pollution events such as wildfires \citep{morawska2018applications, bi2020incorporating, delp2020wildfire}.

PurpleAir data offer dense spatial coverage and minute-level sampling, enabling hourly mapping of ambient PM$_{2.5}$ across California with substantially finer spatial granularity than the EPA reference network \citep{barkjohn2020development}. Leveraging high spatiotemporal resolution data collected from 1015 PurpleAir sensors operating from January to December 2019, this study aims to assess various geostatistical and machine learning techniques in producing hourly PM$_{2.5}$ maps. 

In recent years, considerable effort has been devoted to developing statistical methods that accurately predict PM$_{2.5}$ concentrations on the spatial, temporal, and spatiotemporal domains. Traditional statistical methods like kriging and land use regression have been essential for epidemiological studies to estimate PM${2.5}$ levels. However, these methods often suffer from computational intensity and data availability limitations, respectively \citep{alexeeff2015consequences,  hu2013estimating}. 
\citet{datta2016hierarchical} introduced a hierarchical nearest-neighbor Gaussian process model for large geostatistical datasets, which used a hierarchical structure to reduce the computational complexity of the Gaussian process while maintaining its accuracy. However, it requires spatiotemporal covariance structures and parametric assumptions. Beyond classical frameworks, Bayesian geostatistical models implemented via integrated nested Laplace approximations (INLA) with \textcolor{black}{stochastic partial differential equations} (SPDEs) representations provide scalable inference on Gaussian random fields and full predictive uncertainty, and have been applied to PM$_{2.5}$ mapping in recent work \citep{gong2021multivariate, Chen2023}. Implementing INLA would require a different modeling pipeline (SPDE mesh, priors), so we position it as state-of-the-art context rather than a head-to-head comparator in this study.}

More recently, machine learning algorithms have been explored for air pollution prediction, including random forest,  support vector regression, and deep learning approaches. 
Representative studies combine ground observations, satellite products, and meteorology to improve accuracy at high spatial resolution (e.g., random forest, support vector regression and neural network frameworks), while geostatistical baselines remain competitive under appropriate covariance assumptions \citep{hu2017estimating, mogollon2021support, gupta2009particulate, berrocal2020comparison}. Taken together, the current literature spans two paradigms: probabilistic geostatistical models that explicitly encode spatial dependence and provide predictive uncertainty, and flexible machine learning models that capture nonlinearity but typically yield point predictions.

Despite the variety of approaches available, there is no consensus on which produces the most accurate predictions. Several studies have evaluated traditional geostatistical methods with machine learning algorithms. \cite{requia2019evaluation} compared ordinary kriging, hybrid interpolation, and random forest for estimating concentrations of 10 PM$_{2.5}$ components. They demonstrated that random forest surpassed both kriging and hybrid interpolation methods (empirical Bayesian kriging and land use regression). 
\cite{wang2019nearest} introduced a flexible spatial prediction approach using a nearest neighbor neural network, which excelled over conventional geostatistical methods in handling non-normal data by innovatively utilizing neighboring information. Conversely, \cite{berrocal2020comparison} evaluated universal kriging and downscaler models against machine learning approaches for generating daily national maps of PM$_{2.5}$ concentrations in the United States, noting superior predictive performance in geostatistical methods under the assumption of constant spatial covariance.

Although numerous studies have compared geostatistical and machine learning techniques, they often focus solely on spatial or temporal dependencies, neglecting their combined spatiotemporal dynamics. Leveraging the extensive network of PurpleAir sensors, this study aims to produce hourly maps for PM$_{2.5}$ concentrations in California. We explore the efficacy of both traditional geostatistical methods, such as universal kriging, nearest neighbor Gaussian process and fixed rank kriging, and advanced non-geostatistical approaches, including regression, random forest, support vector machines, and neural networks. Additionally, this study integrates geostatistical methods and nearest neighbor data with machine learning algorithms to enhance prediction accuracy and performance. Methodologically, we evaluate short-range interpolation performance in a dense-sensor setting and quantify gains from hybridizing kriging-based predictors with machine learning models. \textcolor{black}{Practically, the resulting hourly maps and associated uncertainty where available can support regulatory and operational tasks such as hotspot surveillance during wildfire events, AQI gap-filling in areas without reference monitors.}
The remainder of the article is organized as follows. Section 2 describes the motivating dataset. Section 3 outlines the methodological framework. Section 4 presents the data analysis and results. Section 5 concludes with a summary of the findings, limitation of the current work and directions for future work.
\section{Data}
PurpleAir sensors are one of the low-cost particulate matter (PM) sensors that have gained popularity for their affordability, accessibility, and ease of use in monitoring outdoor air quality. These sensors use laser-based optical sensors to count particles by sizes and use the counts to calculate the mass concentrations of PM$_{1.0}$, PM$_{2.5}$, and PM$_{10}$. Each sensor has two Plantower PMS5003 units, labeled as channels A and B, which operate alternately and provide 80-second averaged data values (and, from June 2019, 2-minute averaged values) \citep{barkjohn2020development}. Beyond mass, PurpleAir also records meteorological covariates (temperature in °F, relative humidity (RH) in \%, and barometric pressure in hPa), timestamps, and geolocation (latitude/longitude). Widely adopted by communities, citizen scientists, and researchers, PurpleAir sensors are extensively used for local air quality monitoring, and their data has been utilized in various applications. The substantial increase in sensor counts from 251 in 2017 to 39266 in 2022 \citep{API} as shown in Figure \ref{counts}, reflects a growing interest in understanding air quality at the local level. This increase has important implications for air quality monitoring and decision-making. Figure \ref{fig:PA.real} illustrates the rapid increase in PurpleAir installations in California from 2019 to 2023. This underscores the growing importance of incorporating these sensors into air quality monitoring efforts. With more sensors in operation, a higher density of data points can be obtained, allowing a more comprehensive and detailed assessment of air quality trends and patterns in specific areas. 

\begin{figure}[t!]
     \centering
         \includegraphics[trim = 0cm 0cm 0cm 0.7cm, clip, scale=0.6]{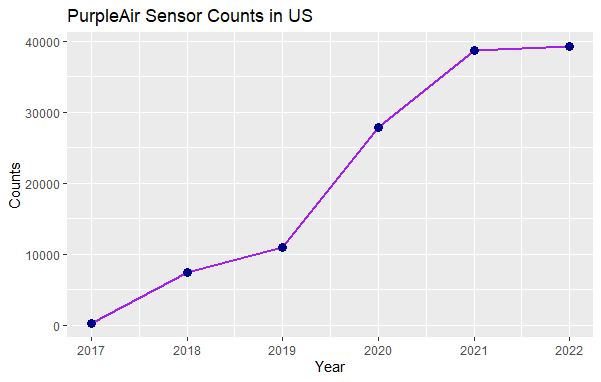}
      \caption{\label{counts} PurpleAir sensors counts from 2017-2022 in US}
\end{figure}

\begin{figure}[H]
 \centering
\includegraphics[scale=0.5]{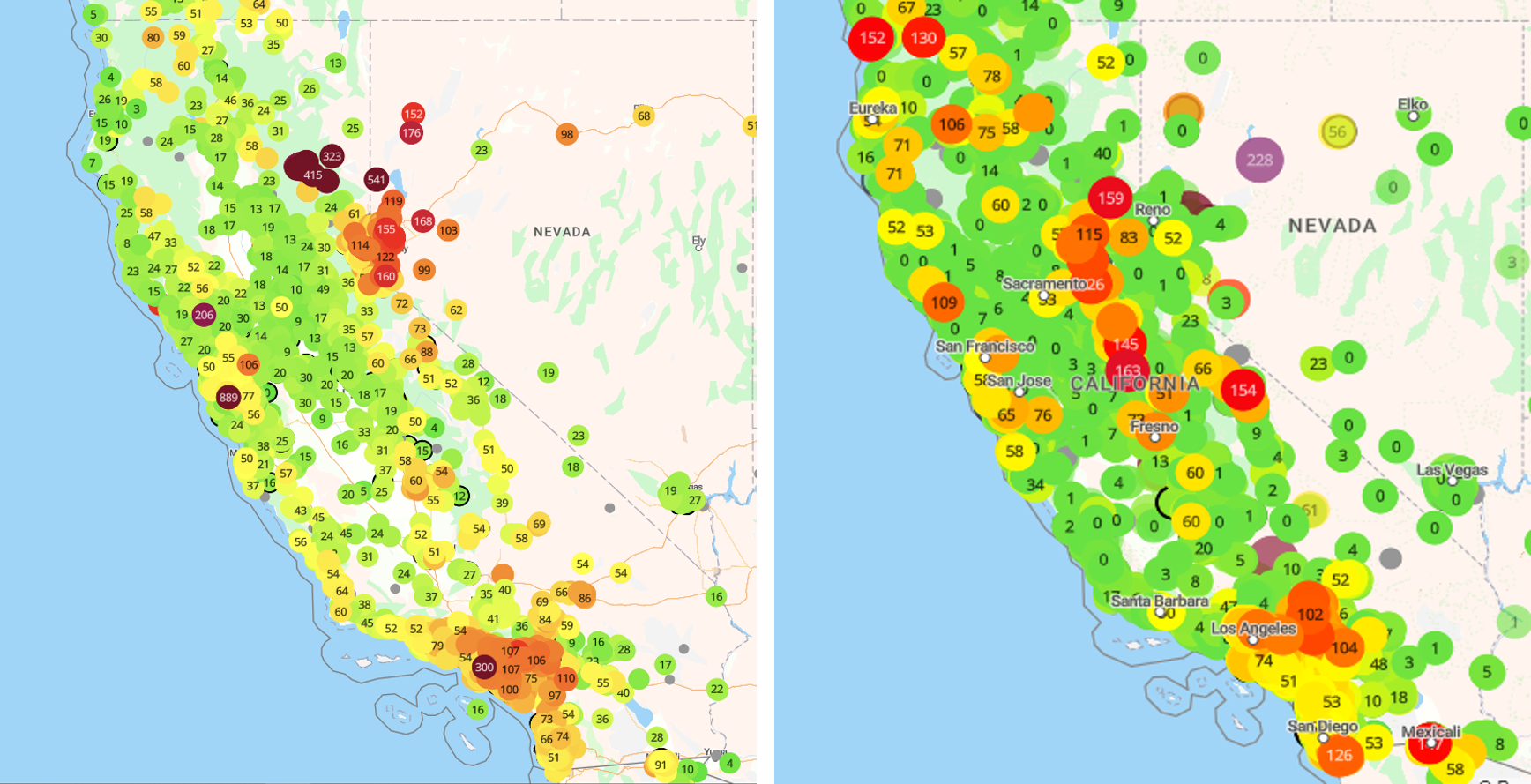}
\caption{\label{fig:PA.real} \textcolor{black}{Real-time map of PM$_{2.5}$ concentration measured by PurpleAir sensors in 2019 (left) and 2023 (right), reproduced from the PurpleAir public visualization interface \citep{purpleair}. Marker colors indicate air quality categories (green) satisfactory; (yellow) acceptable; (orange) members of sensitive groups may be affected; (red) the general public may experience health effects; (purple) increased health risk for everyone. Marker sizes reflect visualization rendering and display resolution in the original interface and do not represent the number or density of installed sensors.}}
\end{figure}

However, it is essential to consider the limitations of PurpleAir sensors, such as potential issues with sensor accuracy, calibration, and data quality. 
As PurpleAir sensor counts continue to rise, it is crucial to ensure that proper quality control measures, calibration, and validation protocols are in place to ensure the reliability and accuracy of the data obtained from these sensors for robust air quality assessments and decision-making. To start with the most robust data, we processed the PurpleAir feeds in three steps after accessing and downloading the data in JSON format: \textbf{(1)} We removed observations with known sentinel/communication codes for temperature (\texttt{2,147,483,447} or $-224\,^{\circ}\mathrm{F}$), excluded physically impossible values ($>1000\,^{\circ}\mathrm{F}$), and restricted temperature to a conservative ambient range ($-58$ to $140\,^{\circ}\mathrm{F}$). We also dropped RH values outside $0$--$100\%$ (including the $255\%$ artifact). After this screening, two of 1015 sensors were fully excluded for having only unreliable temperature readings; all other sensors were retained. The excluded sensors are geographically distant ($\approx 34.1^{\circ}\mathrm{N}, -118^{\circ}$; $37.9^{\circ}\mathrm{N}, -123^{\circ}$). \textbf{(2)} We checked A/B channel consistency and then averaged the 2-minute (or 80-second) measurements to hourly values. \textbf{(3)} We applied the correction formula given by \cite{barkjohn2020development},
\begin{center}
$PM_{2.5corrected}=0.524*PA_{avg}-0.0852*RH+5.72$
\par\end{center}
where PM$_{2.5corrected}$ is the corrected PM$_{2.5}$, $PA_{avg}$ is the hourly averaged PM$_{2.5}$ from channel A and B and $RH$ is hourly relative humidity.
For our study, we collected hourly measurements of PM$_{2.5}$ concentrations from January 2019 through to December 31, 2019, from 1015 PurpleAir sensors in California, of which 1013 remained after QC and were used in modeling. Figure \ref{PA2} illustrates an example of the data.
\begin{figure}[H]
     \centering
         \includegraphics[scale=.7]{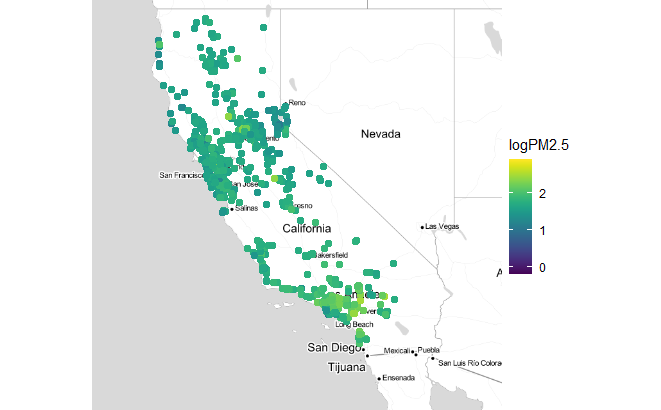}
      \caption{\label{PA2} PM$_{2.5}$ concentration on the log scale on May $18^{th}$, 2019 at 2 pm.}
\end{figure}
\section{Methods}\label{methods}
\subsection{Overview}
Our methodological framework combines three geostatistical models, universal kriging (UK), fixed rank kriging (FRK), and nearest neighbor Gaussian process (NNGP), and several non-geostatistical machine learning models, including random forest (RF), support vector regression(SVR), and neural network (NN). 
These models were selected to represent distinct theoretical paradigms in spatial prediction: UK as a benchmark stationary Gaussian process model, FRK as a reduced-rank approximation suitable for large datasets, and NNGP as a scalable local Gaussian process using conditional independence. 
The machine learning models complement geostatistical methods by relaxing distributional assumptions and capturing nonlinear relationships between predictors. This comparative design enables evaluation across the spectrum from fully parametric spatial models to nonparametric, data-driven algorithms.

\subsection{Geostatistical Methods}
Geostatistical methods are a set of statistical techniques used to analyze spatial/spatiotemporal data and explicitly take into account the spatial/spatiotemporal dependence of the data. These methods are widely used in various fields such as geology, hydrology, agriculture, environmental science, and many others \citep{maliva2016geostatistical}. A typical procedure in geostatistical techniques is to start with modeling the spatial/spatiotemporal correlation structure of the data, which describes how the values at different locations are related to each other in space and over time. Once we estimate the model parameters from the data, we can make predictions for the variable of interest at unsampled locations.

\subsubsection{Universal Kriging}
 Let $Y(s,t)$ be the observed PM$_{2.5}$ ($mg/m^{3}$) from PurpleAir sensors at location $s\in \mathit{\mathscr{\mathcal{D\subset}}\mathbb{R^{\mathrm{2}}}}$ and time $t \in \mathbb{R} $. Then
 \begin{equation}\label{eq:GP}
Y(s,t)=\mu(s,t)+w(s,t)+\epsilon(s,t)
\end{equation}
 where, 
 $\mu(s,t)=\bX(s,t)\bbeta$ is the mean with $\bX(s,t)=[X_{1}(s,t),X_{2}(s,t),\dots,X_{p}(s,t)]^{T}$ are predictors and $\bbeta$ is the $p$-dimensional vector of parameters. $w(s,t)$ represents a mean-zero Gaussian process with a spatiotemporal covariance function. $\epsilon(s,t)$ is the white noise measurement error with  mean $0$ and variance $\sigma_{\epsilon}^{2}$. We use a separable exponential covariance function,
\begin{equation}\label{eq:expcov}
 {\Gamma(\mathbf{h};\tau,\btheta) = {\sigma_s}^2 exp\left({-{||\mathbf{h}||}}/{\rho_s}\right)}{\sigma_t}^2 exp\left({-{|\tau|}}/{\rho_t}\right)
\end{equation}
 where $\bh=s^\prime-s$ is the spatial lag, $\tau=t^\prime-t$ temporal lag and $\btheta=\{\sigma_s,\sigma_t,\rho_s,\rho_t\}$ is the set of parameters associated with this covariance function, where $\sigma_s,\sigma_t$ are the standard deviations for space and time , and $\rho_s,\rho_t$ are spatial and time range parameters respectively. To predict $\bY$ at any unsampled location at $(s_0,t_0)$, we have
 \begin{equation}\label{BLUP3}
    \hat{Y}(s_0,t_0) = \bX(s_0,t_0)^T \hat{\bbeta}_{gls} + c(\btheta)^T  \Sigma_y^{-1}(\btheta)(\bY - \bX \hat{\bbeta}_{gls}),
\end{equation}
where $c(\btheta) = \text{Cov}(Y(s_0,t_0),\bY)$ and $ \text{Var}(\bY)=\Sigma_y(\btheta)$. The uncertainty can be obtained by
\begin{equation}\label{Var3}
    \text{Var}(\hat{Y}(s_0,t_0)) = c_{0,0} - c(\btheta)^T  \Sigma_y^{-1}(\btheta)c(\btheta)+\kappa,
\end{equation}
where $c_{0,0}=\text{Var}(Y(s_0,t_0))$ and 
\begin{center}
$\kappa=\left( \bX(s_0, t_0) - \bX^T \Sigma_y^{-1}(\btheta) c(\btheta) \right)^T \left( \bX^T \Sigma_y^{-1}(\btheta) \bX \right)^{-1} \left( \bX(s_0,t_0) - \bX^T \Sigma_y^{-1}(\btheta) c(\btheta) \right)$.
\end{center}
The above prediction method is called universal kriging. To obtain the prediction in Equation \eqref{BLUP3} and the prediction uncertainty in Equation \eqref{Var3}, we compute the estimates $\hat{\bbeta}_{gls}$ and  $\hat{\btheta}$ to yield the empirical best linear unbiased predictor (EBLUP).
Under the model specification, the random variable $Y(s_0, t_0) | \mathbf{Y}$ representing the value at an unobserved location conditional on the observed data, follows a Gaussian distribution. The mean of this distribution is given by Equation \eqref{BLUP3}, $\hat{Y}(s_0, t_0)$, and its variance is given by Equation \eqref{Var3}, $\text{Var}(\hat{Y}(s_{0},t_{0}))$. Consequently, a 95\% prediction interval for $Y(s_0, t_0)$ is derived directly from this Gaussian predictive distribution and is given by:

\begin{equation}\label{eq:predcover}
     \hat{Y}(s_0,t_0) \pm Z_{0.975} \times \sqrt{\text{Var}(\hat{Y}(s_0,t_0))},
\end{equation}

where $Z_{0.975} \approx 1.96$ is the 97.5th percentile of the standard normal distribution. In this study, we implemented universal kriging using the gstat package in R \citep{graler2016spatio}
\subsubsection{Nearest Neighbor Gaussian Process}
 The nearest neighbor Gaussian process extends Vecchia’s approximation \citep{vecchia1988estimation} to a process using conditional independence given information from neighboring locations \citep{datta2016hierarchical}. For notation purposes, let
\begin{equation}\label{Y3}
    \bY = \begin{pmatrix} Y(\bs_{1}, t_{1}) , Y(\bs_{2}, t_{1}) , \cdots ,Y(\bs_{m}, t_{1}) , \cdots , Y(\bs_{1}; t_{T}) , \cdots , Y(\bs_{m},t_{T}) \end{pmatrix}^\prime.
\end{equation}
The joint density of $\bY$ can be written as the product of conditional densities, as
\begin{equation}\label{condY}
    f(\bY) = \prod\limits_{i=1}^N f(Y_{i}|Y_{1},\dots,Y_{i-1})
\end{equation}
The idea for this method is to approximate $f(\bY)$ given in \eqref{condY}  by
\begin{equation}\label{eq:NN1}
f(\bY)\approx \Tilde{f}(\bY) =\prod\limits_{i=1}^{N} f(\bY_{i}|\bY_{\mathcal{N}_{i}})
\end{equation}
where 
$\mathcal{N}_{i}$ is the neighboring set for $\bY_{i}$ and $\bY_{\mathcal{N}_{i}}$ are the observations in this set. \cite{vecchia1988estimation} demonstrated that approximating the full conditional with a subset of $m$ nearest neighbors provides an excellent approximation in the case of spatial data and \cite{datta2016nonseparable} extended it to spatiotemporal data. In this study, we construct the $m$ nearest neighbors using simple neighbor selection suggested by \cite{datta2016nonseparable} and select the $m$ nearest spatial neighbor with temporal lag 1. The NNGP is implemented with GpGp package in R \citep{GpGp}. GpGp relies on ordering of the Gaussian process observations where the conditional distribution only conditions on a small subset of previous observations in the ordering. This results a sparse Cholesky factor of the precision matrix. After the estimation step, to predict at any unsampled location $(s_0,t_0)$, 
\begin{equation}
    \hat{Y}(s_0,t_0) = \bX(s_0,t_0)^T \hat{\bbeta} + C_{s_0,{\mathcal{N}_{0}}}C_{{{\mathcal{N}_{0}}}}^{-1} (\bY_{\mathcal{N}_{0}} - \bX(s_0,t_0)^T \hat{\bbeta}),
\end{equation}
where $C_{s_0,{\mathcal{N}_{0}}}=\text{Cov}(Y(s_0,t_0),\bY_{\mathcal{N}_{0}})$ and $C_{{{\mathcal{N}_{0}}}}= \text{Var}(\bY_{\mathcal{N}_{0}})$.

\subsubsection{Fixed Rank Kriging}
Fixed rank kriging is a spatial/spatiotemporal interpolation technique that combines the strengths of kriging with a reduced rank approximation to address the computational challenges of large spatial datasets \citep{cressie2008fixed}. The Gaussian process here is modeled similarly to \eqref{eq:GP} except that $\bw(s,t)$ is approximated by a linear combination of $\mathbf{K}$ basis functions $\bf {\mathbf{\phi}(s,t)=(\phi_1(s,t),\phi_2(s,t),\dots,\phi_\mathbf{K}(s,t))}$ and the basis function coefficients
$\bw^{*}=(w_1^{*},w_2^{*},\dots,w_\mathbf{K}^{*})$\citep{wikle2019spatio, heaton2019case}:

 \begin{equation}\label{eq:FRK1}
\bw(s,t) \approx \Tilde{\bw}(s,t)=\sum\limits_{i=1}^{\mathbf{K}} \phi_i(s,t) w_\mathbf{i}^{*}.
\end{equation}
This approximation reduces computational demands by ensuring that all estimations and predictions involve matrices of size $\mathbf{K} \times \mathbf{K}$, where $\mathbf{K} \ll \mathbf{N}$ and $\mathbf{N}$ is the total number of observations. Moreover, $\bf \mathbf{\phi}(s,t)$ can be composed at $R$ different resolutions, thus 
\begin{equation}\label{eq:FRK2}
 \Tilde{\bw}(s,t)=\sum\limits_{r=1}^{\mathbf{R}}\sum\limits_{i=1}^{\mathbf{K_r}} \phi_{ri}(s,t) w_\mathbf{ri}^{*},
\end{equation}
where $\phi_{ri}(s,t)$ is the $i$th spatiotemporal basis function at the $r$th resolution with basis function coefficients $w_\mathbf{ri}^{*}$ and the total number of basis functions is given by $\mathbf{K}=\sum\limits_{r=1}^{\mathbf{R}}\mathbf{K_r}$. The coefficients $\bw^{*}=(w_\mathbf{ri}^{*}: r=1,\dots,R, i=1,\dots,\mathbf{K_r})$ can be modeled as spatial, temporal or spatiotemporal varying \citep{wikle2019spatio}. Here we consider the case where $\bw^{*}$ is spatial varying, thus $\bw^{*}\sim \text{MVN} (0,\Sigma_{\bw^{*}}(\btheta)) $ and $\Sigma_{\bw^{*}}(\btheta)$ is chosen to be exponential covariance function,  with $\btheta =(\sigma_s,\rho_s, \sigma_t =1, \rho_t=0)$ in equation \eqref{eq:expcov}.
In this study, we construct $\{\phi_i(s,t),i=1,\dots,\mathbf{K}\}$ by taking the tensor product of a spatial basis function with a temporal basis function \citep{wikle2019spatio}. Thus, the spatiotemporal basis functions $\mathbf{\phi}(s,t)=\{\phi_{st,u}:u=1,\dots,r_{s}r_{t}\}=\{\phi_{p}(s)\psi_{q}(t):p=1,\dots,r_{s};q=1,\dots,r_{t}\}$, where $r_{s}r_{t}$ are sets of spatial and temporal basis functions respectively. In this study, we choose bisquare basis functions for both spatial and temporal, we construct ${\phi_{ri}(s,t)}$ as tensor products of bisquare spatial and temporal bases placed with a space-filling design at $R=2$ spatial resolutions and a single temporal resolution. Spatial radii bracket the empirical correlation range; temporal knots target seasonal variability. The resulting
$K=\sum_{r=1}^{R}K_r=1{,}600$ ($80$ spatial $\times$ $20$ temporal) yields a reduced-rank FRK with $K\ll N$. We conducted a sensitivity analysis over $K\in{800,1200,1600,2000}$, one vs.\ two spatial resolutions, and temporal knots ${10,20,30}$ using rolling time-blocked cross-validation. Predictive accuracy, \textcolor{black}{measured by the root mean squared error (RMSE), mean absolute deviation (MAD)}, and interval calibration varied only marginally across settings, with the two-resolutions, $K=1{,}600$, 20-knot configuration offering the best accuracy–cost balance.
With parameter estimates $\hat{\bbeta}$, the FRK predictor at a new location $(s_0,t_0)$ is given by 
  \begin{equation}\label{eq:FRK.pred}
 \hat{Y}(s_0,t_0)=\bX(s_0,t_0)\hat{\bbeta}^\prime+ c(s_0,t_0)^T  \Sigma_y^{-1}(\hat{\btheta})(\bY - \bX \hat{\bbeta}),
\end{equation}
where $\bY$ as defined in \eqref{Y3}, $c(s_0,t_0) = \text{Cov}(Y(s_0,t_0),\bY)$ and $ \text{Var}(\bY)=\Sigma_y(\btheta)$. The uncertainty is given by
\begin{equation}\label{FRKVar}
    \text{Var}(\hat{Y}(s_0,t_0)) = \mathbf{\phi}(s_0,t_0)^T \Sigma_{\bw^{*}}(\hat{\btheta})\mathbf{\phi}(s_0,t_0)+ \sigma_{\epsilon}^{2}-c(s_0,t_0)^T \Sigma_y^{-1}(\hat{\btheta}) c(s_0,t_0).
\end{equation}
We implemented this method using the FRK package in R \citep{zammit2021frk}. 
\subsection{Non-geostatistical Methods}
In this section we will present methods that do not account for the spatiotemporal dependence in the data. We present the commonly used regression method and several machine learning algorithms.
In recent years, machine learning algorithms have shown promise in modeling environmental data. Unlike traditional statistical models, machine learning algorithms don't rely on assumptions about the underlying distribution of the data and can identify complex patterns in the data and learn from them to make accurate predictions \citep{breiman2001statistical, reid2015spatiotemporal, di2016assessing}. Furthermore, these algorithms can handle high-dimensional data and model complex non-linear relationships between variables \citep{athmaja2017survey}. For the machine learning algorithm, we will discuss regression, random forest, support vector regression, and neural network with a brief explanation for each algorithm. 
\subsubsection{Regression}
Consider a regression model where we assume that we will account for all spatiotemporal dependence in the predictors. The typical regression model is given by
 \begin{equation}\label{eq:Reg}
Y(s,t)=\bmu(s,t)+\bepsilon(s,t),
\end{equation}
where, $\bmu(s,t)=\bX(s,t)\bbeta$ is the mean with $\bX(s,t)=[X_{1}(s,t),X_{2}(s,t),\dots,X_{p}(s,t)]^{T}$ are predictors and $\bbeta$ is the $p$-dimensional vector of parameters and \bepsilon(s,t) is the white noise measurement error with mean $0$ and variance $\sigma_{\bepsilon}^{2}$. In this model, we will choose $\bX(s,t)$ in terms of the nearest neighbor criteria. Although this model is simple to implement, it accounts for model error and allows us to obtain prediction error variance.

\subsubsection{Random Forest}
Random forest regression is a supervised learning algorithm that uses an ensemble of decision tree learning methods for regression. 

First, the main idea of decision trees is to make a tree that predicts a regression surface $\hat{f}(X|\hat{c})=\sum_{i=1}^M \hat{c}_i I(X\in R_i)$ for a partition of regions $\{R_i:i=1,\ldots,M\}$ and $\hat{c}_i \in \mathbb{R}$ for $i=1,\ldots,M$. In each decision tree, a random sample of
$m$ predictors is chosen as split candidates from the full set of $p$ predictors \citep{breiman2001statistical}. The algorithm uses the method of least squares to minimize $RSS(c)=\sum_{i=1}^N (y_i-f(x_i|c))^2$. It results in $\hat{c}_m=\frac{1}{N_m}\sum_{i:x_i\in R_m}y_i$
where $N_m=|\{i:x_i\in R_m\}|$. The $\{R_i:i=1,\ldots,M\}$ is found using a greedy algorithm to grow the regression tree top-down. Decision trees method is not always a strong machine learning method and can be improved using ensembles. 

Random forest is a non-parametric ensemble machine learning algorithm that requires the selection of two key parameters:  the number of predictors in the random subset of each node ($m$) as the default value and the number of decision trees in the forest ($n$). This algorithm constructs multiple decision trees using bootstrapped training samples. At each node, a random sample of $m$ features from the total $p$ features is selected, and the algorithm identifies the optimal feature for creating a split. Predictions from the $n$ trees are then aggregated using the mean to determine the final output. The error rate is assessed using predictions of out-of-bag samples \textcolor{black}{\citep{hastie2009elements}}. In this study, we used 500 trees to minimize the out-of-bag error, confirmed through five-fold cross-validation. Prediction coverage was evaluated by calculating the 2.5\%  and 97.5\% quantiles across all 500 trees. This was implemented using the randomForest package in R \citep{liaw2002classification}.

\subsubsection{Support Vector Regression}
The groundwork for support vector machine was provided by \cite{vapnik1974theory} and \cite{vapnik1999nature}. A regression version of that, named support vector regression, was introduced in \cite{Drucker1997}. Based on selected kernel $k(x,y)$, a predictive model with input features $\mathbf{x}$ is given by
 \begin{equation}\label{eq:SVR1}
F(x|w)=\sum_{i=1}^N (\alpha_i-\alpha_i^*)K(v_i,\mathbf{x})
\end{equation}
for vectors $(\alpha_1,\ldots,\alpha_N)$ and $(\alpha_1^*,\ldots,\alpha_N^*)$ and support vector $v_i$. Here we use the radial kernel $k(x,y)=\exp(-\dfrac{\sum_i (x_i-y_i)^2}{\gamma})$. The set of parameters is given by $w$. The primal objective function is
 \begin{equation}\label{eq:SVR2}
\lambda\sum_{i=1}^N \ell(y_i-F(x_i|w))+\lVert w \rVert_2^2.
\end{equation}
Note that the regularization parameter $\lambda$ is introduced in front of the loss function and not the $L_2-$regularization term. The best value of $\lambda$ is to be found using cross-validation.
Either $\alpha_i$ or $\alpha_i^*$ (but not both) will be non-zero based on the location of the observed point above or below the $\epsilon-$tube, respectively. Both are zero if the point falls inside the tube. The $\epsilon-$tube is induced by the $\epsilon-$sensitive loss function defined so that it has a value of zero if the predicted value is within the tube. 

The dual maximization problem is to
\begin{equation}\label{eq:SVR2.1}
\text{max$_{\alpha,\alpha^*}$ } - \frac12 \sum_{i=1}^N \sum_{j=1}^N K(v_i,v_j) (\alpha_i-\alpha_i^*)(\alpha_j-\alpha_j^*) -\sum_{i=1}^N y_i (\alpha_i-\alpha_i^*) -\epsilon \sum_{i=1}^N (\alpha_i+\alpha_i^*)
\end{equation}
subject to
 \begin{equation}\label{eq:SVR3}
\sum_{i=1}^N \alpha_i=\sum_{i=1}^N \alpha_{i}^*,\; 0\leq \alpha_i,\alpha_i^* \leq \lambda
\end{equation}
for $i=1,2,\ldots,N.$

The dual maximization problem is equivalent to a quadratic programming problem 
 \begin{equation}\label{eq:SVR4}
\text{min$_\mathbf{\beta}$ }\frac12 \mathbf{\beta}' Q \mathbf{\beta} + \mathbf{c}' \mathbf{\beta}
\end{equation}
subject to 
 \begin{equation}\label{eq:SVR5}
\sum_{i=1}^N \beta_i=\sum_{i=1}^N \beta_{i+N},\; 0\leq \beta_i \leq \lambda
\end{equation}
for $i=1,2,\ldots,N$, where $\beta_i=\alpha_i^*$ and $\beta_{i+N}=\alpha_i$ for $i=1,2,\ldots,N$ and $c_i=\epsilon-y_i$ and $c_i=\epsilon+y_i$ for $i=1,2,\ldots,N$ while the block matrix $Q=\begin{bmatrix}
D & -D\\
-D & D
\end{bmatrix}$ with $D_{i,j}=K(v_i,v_j)$.
We implemented SVR using the e1071 package in R \citep{dimitriadou2008misc}.

\subsubsection{Neural Network}
Neural network algorithms are intended to mimic human brain learning processes; see Figure \ref{fig:ANN}. A neural network possesses input data from an input layer through a linear map ${\mathbf{L}}_1: X\mapsto X_{L_1}$ to the first hidden layer. The linear map ${\mathbf{L}}_1(X)=\mathbf{W}^{(1)}X+b^{(1)}$, where $\mathbf{W}^{(1)}$ is a $M_1\times p$ matrix and $b^{(1)}$ is $M_1\times 1$ bias vector. An activation function $\sigma_1:X_{L_1}\mapsto Z^{(1)}$ activates each of the $M_1$ neurons in the first layer. The same process, using linear maps $\mathbf{L}_i$ and activation functions $\sigma_i$, takes place between the $i-1^{th}$ and $i^{th}$ hidden layers with $M_{i-1}$ and $M_i$ neurons, respectively, for $i=1,\ldots,N$. The linear maps ${\mathbf{L}}_i(X)=\mathbf{W}^{(i)}X+b^{(i)}$, where $\mathbf{W}^{(i)}$ is a $M_i\times M_{i-1}$ matrix and $b^{(i)}$ is $M_i\times 1$ bias vector. A target linear map $\mathbf{T}: Z^{(N)}\mapsto T$ precedes the output function $g_k:T \mapsto Y_K$ at the output layer. In overall, it is a composition ($\bigcirc$) of linear and activation functions $$Y_k=g_k(\mathbf{T}(\bigcirc_{i=1}^N\sigma_i(\mathbf{L}_i(Z^{(i-1)}))))=:f_k(X|\Theta)$$ with $Z^{(0)}:=X$ and $f_k:X \mapsto Y_k$ is a nonlinear transformation of $X$ into $Y$. In addition, $\Theta$ is a vector of all parameters of the neural network that include the weights/entries in the matrices and the biases of the linear maps and the parameters of the activation and output functions. 
An optimization algorithm is used to minimize a loss function $\ell(\Theta)=\frac{1}{K}\sum_{i=1}^K \ell_i(y_k,f_k(X|\Theta))$. The loss function $\ell_i$ is usually the mean squared error $\ell_i(x,y)=\sum_{j=1}^n (x_j-y_j)^2$ in case of regression problems and cross-entropy $\ell_i(x,y)=-\sum_{j=1}^n x_j\log(y_j)$ in case of classification problems. An $L_p-$ regularization is also used in which the objective function is $\ell(\Theta)+\lambda \lVert \Theta \rVert_p^p$ for $p=1,2$. The hyperparameter is tuned using cross-validation. The most common algorithm to find the optimal solution $\Theta$ is using a back-propagation algorithm with the stochastic gradient descent method.

There are different types of activation functions that can be used to activate neurons
    \begin{itemize}
    \item Identity: $\sigma(x)=x$
\item Sigmoid: $\sigma(x)=S(x)=\frac{1}{1+e^{-x}}$
\item Hyperbolic tangent: $\sigma(x)=tanh(x)=\frac{e^x-e^{-x}}{e^x+e^{-x}}$
    \item Rectified Linear Unit (ReLU): $\sigma(x)=ReLU(x)=max(x,0)$
        \item Rectified softplus: $\sigma(x)=ReSP(x)=\log(1+e^{x})$
    \end{itemize}
There are two main output functions $g_k$:
    \begin{itemize}
    \item $k$th element ``identity" function: $g_k(T)=T_k$, which is used for regression problems
        \item Softmax function: $g_k(T)=\frac{e^{T_k}}{\sum_{\ell=1}^K e^{T_\ell}}$, which is used for classification problems.
    \end{itemize}
\begin{figure}[t]
 \centering
\includegraphics[scale=0.45]{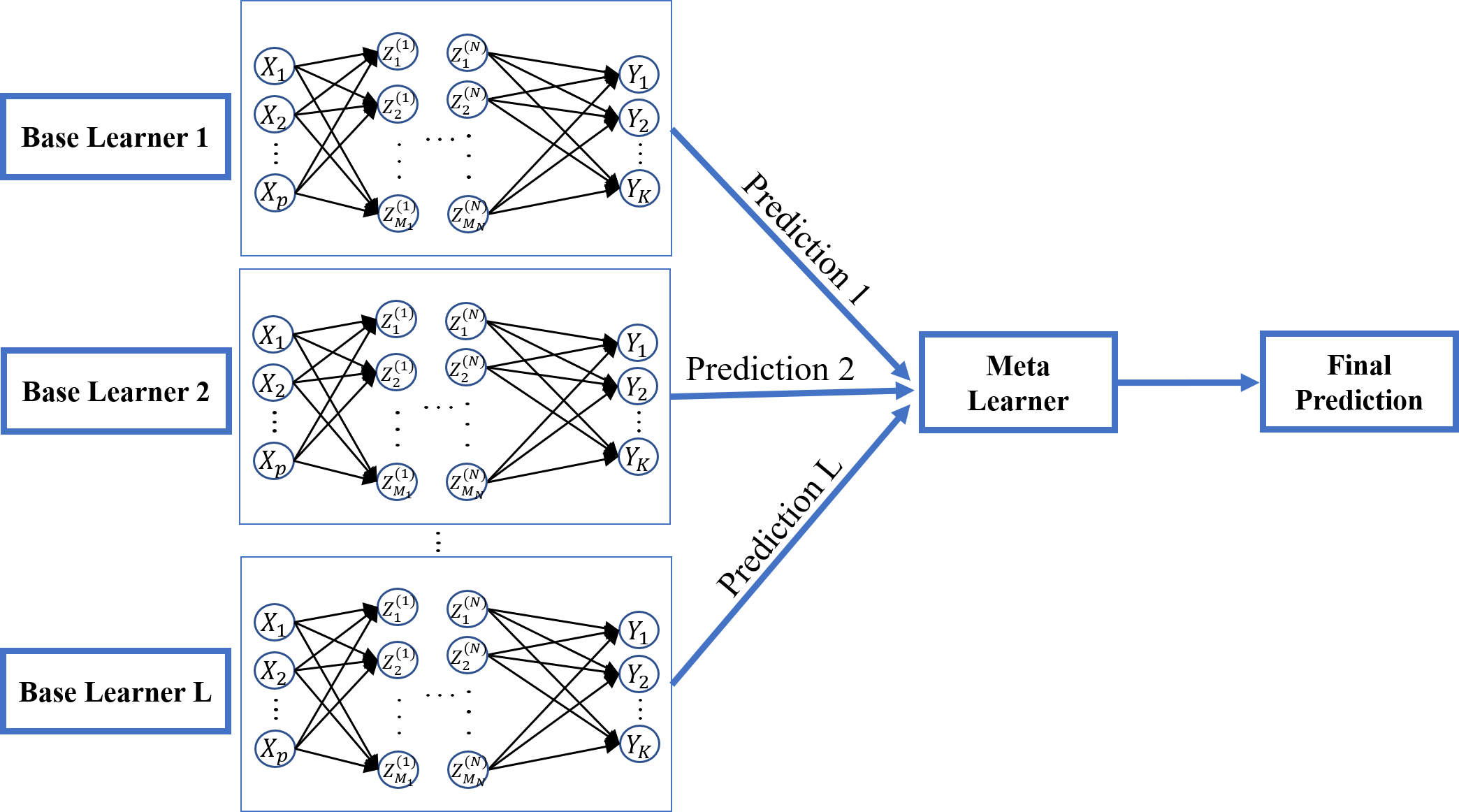}
\caption{\label{fig:ANN}Illustration of a Multi-layer Neural Network with $N$ hidden layers.}
\end{figure}
Ensemble neural networks (ENN) are a type of ensemble method that use multiple neural network models, each with their own architectures, hyperparameters, and training data, to collectively make predictions. The idea behind ENNs is that by leveraging the complementary strengths and weaknesses of multiple models, we can achieve better performance than any individual model alone. There are several ways to construct an ENN. One common approach is to train several independent neural networks, called base models or base learners, on the training data and then combine their output through some form of averaging, in the case of regression, or voting, in classification problems \citep{hansen1990neural}. Another approach is to use a single architecture and hyperparameters, but vary the initial weights and biases or the order in which the data are presented during training \citep{hastie2009elements}. In this study, we constructed two neural networks, and their predictions are combined to make a final omnibus prediction. Combining the predictions from the two neural networks can be done either by averaging or by assigning weights to each prediction according to the importance of the model, called stacking. The idea behind stacking is to use the predictions of multiple base models to train a higher-level ``meta-model" that can make more accurate predictions than any of the individual models alone. In our implementation, the ENN uses two feed-forward neural networks (Base Learner 1 (NN1), Base Learner 2 (NN2)) trained by back-propagation with a stochastic gradient–based optimizer (Adam). We implemented the ENN using Keras \citep{chollet2015keras} and Sklearn \citep{pedregosa2011scikit} in Python \citep{van1995python}. NN1 is a wide–shallow network with hidden layers (512, 128), and NN2 is a deeper–narrower network with hidden layers (256, 128, 64). All hidden layers use ReLU activations with dropout and $\ell_2$ weight decay and the loss is the root mean squared error. The hyperparameters were tuned through a cross-validation grid search: learning rate $\in\{10^{-4},3{\times}10^{-4},10^{-3}\}$, batch size $\in\{256,512,1024\}$, dropout $\in[0.1,0.4]$, and penalty $\ell_2$ $\lambda\in\{10^{-4},10^{-3}\}$. The models were trained for up to 200 epochs with early stopping and reduction of learning-rate on the plateau; the configuration that minimizes cross-validation RMSE was selected. To combine base learners we used stacking consistent with \cite{hansen1990neural}: out-of-fold predictions from NN1 and NN2 formed two meta-features and a linear meta-learner (ridge regression with non-negativity and sum-to-one normalization of the weights) learned $(w_1,w_2)$, yielding the final predictor $\widehat{y}_{\mathrm{ENN}}(x)=w_1\,\widehat{y}_{\mathrm{NN1}}(x)+w_2\,\widehat{y}_{\mathrm{NN2}}(x)$ with $w_1,w_2\ge 0$, $w_1{+}w_2=1$. After tuning, each base model was refit on the full training set with its selected hyperparameters, and the meta-learner was applied to their predictions; random seeds were fixed for reproducibility, and we report per-fold and aggregated RMSE for NN1, NN2, and the stacked ENN alongside the learned $(w_1,w_2)$.

\section{Data Analysis}
\subsection{Feature Selection and Comparative Framework}
Based on the set of features associated with each model and whether the model is considered to be a geostatistical model or non-geostatistical model, we divided all models into four groups. Group 1 contains geostatistical models: universal kriging (UK), nearest neighbor Gaussian process (NNGP), and fixed rank kriging (FRK). Group 2 contains non-geostatistical methods, namely regression (Reg), random forest (RF), support vector regression (SVR), and ensemble neural network (ENN). The set of features ($p$) for groups 1 and 2 will be the longitude and latitude for each location. For group 3, we will add the set of the nearest neighbor observations (NNO) for each location and their longitude and latitude as features for Reg, RF, SVR, and ENN.  Finally, for group 4 we will add the kriging that we obtained from NNGP to be an additional feature in the fitting of Reg, RF, SVR, and ENN. 

We intentionally limited the predictors to spatial coordinates, nearest neighbor observations, and kriging-based features rather than including additional meteorological or land-use variables. This restriction allows a controlled comparison of model architectures by removing confounding effects of covariate variability. The goal is to isolate the capacity of each modeling approach to capture spatial dependence and short-term variability, rather than to build a comprehensive environmental model.
While inclusion of meteorological covariates (e.g., temperature, humidity, wind speed) may further improve accuracy, that direction is beyond the scope of this comparative methodological study.

\subsection{Performance Evaluation}
We evaluate the predictive performance across models via five-fold cross-validation by randomly 
sampling, without replacement, 20\% of the PA sensors to be in the test set in 
each of the five folds. Within each fold, we train each of the models on 
the data from the remaining 80\% of the PA sensors, and we generate predictions at the test sites.
This design provides an assessment of short-range predictive accuracy, which is most relevant for the dense PurpleAir network where sensors are geographically close and interpolation is of greater interest than long-range extrapolation.  Although spatially blocked validation can reduce spatial dependence between folds, it tends to over-penalize performance when data exhibit fine-scale spatial clustering. Our random spatial splitting thus offers a balanced and fair framework for comparing relative model performance at local scales. We emphasize that the purpose of this validation scheme is to evaluate interpolation ability within the existing sensor network, rather than long-range extrapolation across unsampled regions.

Under this framework, two prediction tasks can be defined. The first is spatial interpolation, estimating PM$_{2.5}$ concentrations at locations without PA sensors, which is the primary focus of this study.  The second is temporal forecasting of PM$_{2.5}$ at existing sensor sites, which is beyond the scope of the current analysis.

Model performance was assessed using five metrics: root mean squared error (RMSE) in  \eqref{RMSE2}, symmetric mean absolute percentage error (SMAPE) in \eqref{SMAPE2}, mean absolute deviation (MAD) in \eqref{MAD}, correlation (Cor) between predicted and observed PM$_{2.5}$ concentrations, and 95\% prediction coverage. For geostatistical methods, we constructed a 95\% prediction interval as illustrated in \eqref{eq:predcover}.
\begin{equation} \label{RMSE2}
    RMSE =\sqrt{\dfrac{1}{n}\sum\limits_{i=1}^{n} ({Y_{i}-\hat{Y_{i}}})^{2}};
\end{equation}
\begin{equation} \label{SMAPE2}
    SMAPE =\dfrac{100\%}{n}\sum\limits_{i=1}^{n} \dfrac{\abs{Y_{i}-\hat{Y_{i}}}}{\abs{Y_{i}}+\abs{\hat{Y}_{i}}};
\end{equation}

\begin{equation} \label{MAD}
    MAD =\dfrac{1}{n}\sum\limits_{i=1}^{n} {\abs{Y_{i}-\hat{Y_{i}}}},
\end{equation}
where $n$ denotes the number of test samples, $Y_{i}$ represents the actual PM$_{2.5}$ concentration, and $\hat{Y}{i}$ denotes the corresponding predicted value. These metrics are essential to assess the accuracy of our predictive models.

All model training and evaluation were conducted using the same training–test splits and metrics for consistency. 
The hyperparameters of the machine learning models were tuned using five-fold cross-validation, while the geostatistical parameters were estimated by maximum likelihood. \textcolor{black}{After cross-validation was used to assess predictive performance, a final version of each model was fit using the full dataset with the same specifications and used for all subsequent analyses.
}
To evaluate model adequacy beyond predictive accuracy, we assessed whether fitted models adequately removed spatial dependence from the residuals. Specifically, residuals at held-out test locations were examined using global Moran’s I statistics, which are standard diagnostics for detecting remaining spatial autocorrelation. \textcolor{black}{All reported performance metrics and residual diagnostics in this manuscript should therefore be interpreted as measures of within-area interpolation performance rather than as assessments of generalization to spatially independent regions.\\}

\subsection{Results}
In Group~1 (geostatistical baselines), UK was the strongest (RMSE = 0.3730, SMAPE = 7.802\%, MAD = 0.2630, $\mathrm{Cor}=0.8701$) with 95\% coverage of 93\% but very high computation (3{,}600 min); NNGP was far faster (5 min) with modestly higher error (RMSE = 0.4076; coverage = 89\%), and FRK yielded RMSE = 0.4227 (420 min; coverage = 85\%). Non-geostatistical models alone (Group~2) underperformed these baselines (e.g., SVR RMSE = 0.4186, $\mathrm{Cor}=0.6333$). Augmenting machine learning models with nearest-neighbor observations (Group~3) produced large gains at modest cost, for example, SVR+NNO improved to RMSE = 0.1552, SMAPE = 2.740\%, MAD = 0.0920, $\mathrm{Cor}=0.9249$ in $\sim$5.11 min. Adding a single NNGP-kriging predictor (Group~4) gave the best overall performance: SVR+NNO+Krig achieved RMSE = 0.0819, SMAPE = 1.330\%, MAD = 0.0449, $\mathrm{Cor}=0.9792$ in $\sim$10.93 min; \textit{Reg}+NNO+Krig and \textit{RF}+NNO+Krig delivered similar accuracy (RMSE = 0.0826 and 0.0989, respectively) with well-calibrated uncertainty (95\% coverage = 97\% and 96\%). Overall, relative to UK, the top hybrid (SVR+NNO+Krig) reduced RMSE by $\approx$78\% (0.3730 to 0.0819) while remaining orders of magnitude faster in computation (about 11 vs.\ 3{,}600 min), illustrating a favorable accuracy–compute trade-off for the hybrid approach.

\begin{figure}[H]
 \centering
\includegraphics[scale=0.45]{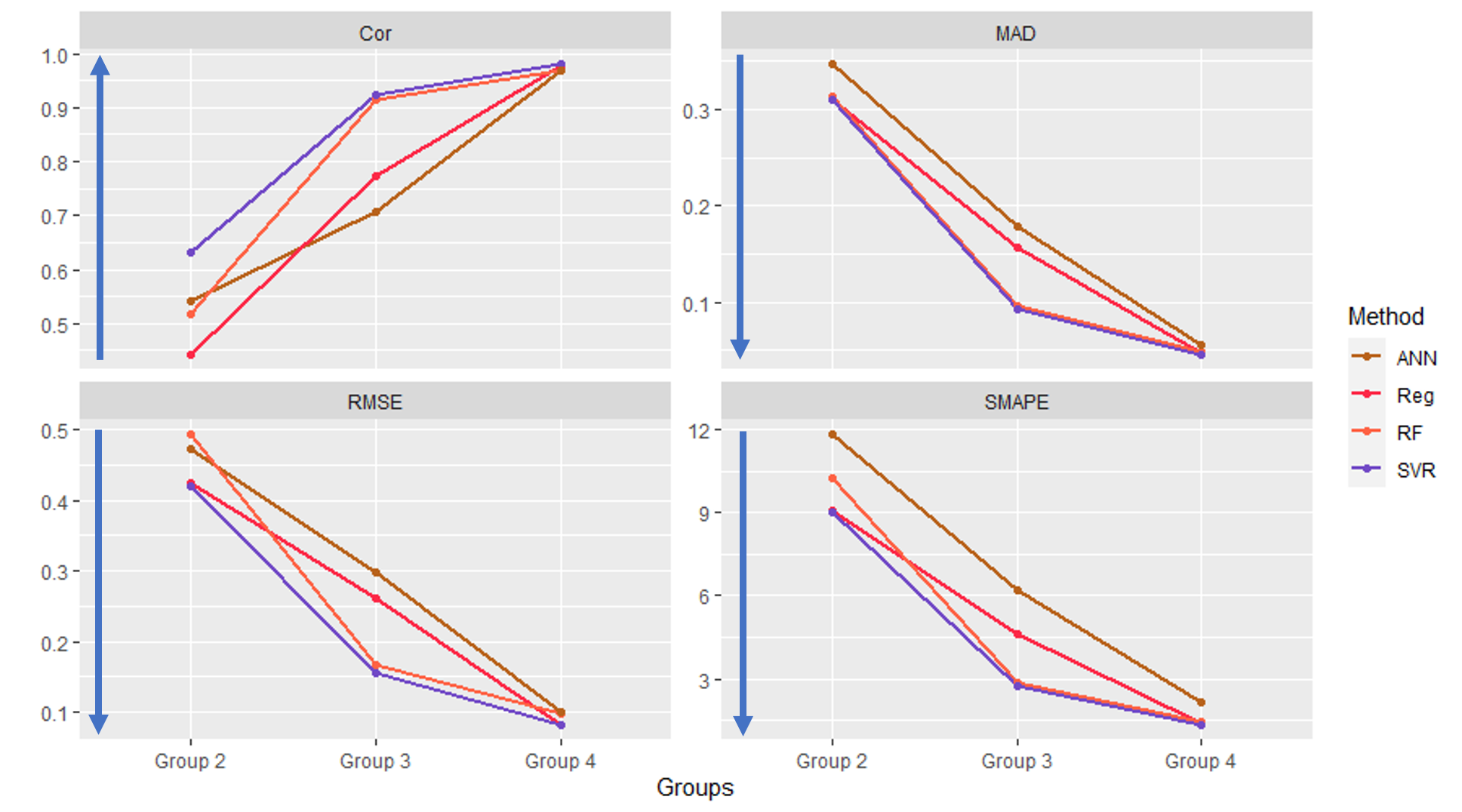}
\caption{\label{fig:metric}A comparison between all models in the Non-geostatistical groups with respect to root mean square error (RMSE), symmetric mean absolute percentage error (SMAPE), mean absolute deviation (MAD), and the correlation (Cor) between observed and predicted values.}
\end{figure}
Figure \ref{fig:metric} illustrates how altering the set of predictors can improve RMSE, SMAPE, and Cor in groups 2,3 and 4. It is worth mentioning that, although the RF algorithm comes in second place after SVR, it offers a unique advantage: it allows for the construction of 95\% empirical prediction coverage. It is important to note that not all models considered in this study provide uncertainty quantification (UQ) in a comparable form. Geostatistical models (UK, FRK, NNGP) yield analytical predictive variances through their Gaussian process formulations, and Random Forest allows empirical uncertainty estimation from the variability among trees. However, Support Vector Regression (SVR) and Ensemble Neural Network (ENN) models are deterministic point predictors that do not natively produce predictive variance. Obtaining UQ for these models would require substantial modification, such as Bayesian SVR, quantile regression, or Bayesian neural networks which falls outside the methodological comparison framework of this paper. For this reason, we report 95\% empirical coverage only for models where valid uncertainty estimates are available.

The rightmost column in Table \ref{table:metrics} also reports the end-to-end elapsed time for each model, including feature construction (adding nearest-neighbor observations, NNO, for Group 3; and the NNGP kriging feature for Group 4), model fitting, and prediction/evaluation. Relative to Group 2, adding NNO features increased the runtime only modestly, for example, SVR from 1.51\,min to 5.11\,min and RF from 3.57\,min to 6.01\,min, while markedly improving accuracy (SVR RMSE from 0.4186 to 0.1552; RF RMSE from 0.4927 to 0.1674). Incorporating the NNGP kriging surrogate in Group 4 produced a further but still modest time increase, for example, SVR+NNO from 5.11\,min to 10.93\,min, accompanied by additional accuracy gains (SVR RMSE from 0.1552 to 0.0819) and improved calibration for models reporting coverage. Overall, these results indicate a favorable computation-accuracy trade-off: small increases in end-to-end time, fully accounting for upstream feature construction costs, yield substantial predictive benefits.
\textcolor{black}{To assess model adequacy and remaining spatial dependence, we evaluated residual spatial autocorrelation for each fitted model using both global Moran’s $I$ statistics. Residuals were computed at held-out test locations as $e_i = Y_i - \hat{Y}_i$. For the geostatistical models that explicitly account for spatial dependence—Universal Kriging (UK), Fixed Rank Kriging (FRK), and NNGP—the residual Moran’s $I$ values were close to zero (UK: $I = -0.069$, FRK: $I = 0.021$, NNGP: $I = -0.084$) and statistically non-significant (all $p \ge 0.129$), indicating no detectable residual spatial autocorrelation. For machine-learning models augmented with spatial information through kriging predictions and nearest-neighbor operators, residual spatial dependence was also effectively mitigated. In particular, the hybrid SVR+NNO+Krig, RF+NNO+Krig, Reg+NNO+Krig, and ENN+NNO+Krig models all yielded small Moran’s $I$ values (ranging from $-0.082$ to $0.044$) with non-significant $p$-values ($p \ge 0.215$), indicating no detectable residual spatial autocorrelation at the global scale. Consistent with these results, empirical semivariograms of the residuals (shown in the Appendix) do not exhibit a clear increasing trend with distance or a well-defined range, suggesting no detectable large-scale residual spatial dependence at the spatial scales considered. Together, these diagnostics indicate that both the explicitly spatial models and the proposed hybrid learning frameworks effectively mitigate residual spatial dependence. As a result, model performance can be interpreted without confounding effects from unmodeled spatial structure at the spatial scales considered.
}

To generate an hourly map of PM$_{2.5}$ concentrations in California, we constructed a  $50\times 50$ grid over a 24-hour period. Predictions were made using the highest performing model from each group: UK from group 1, SVR from group 2, SVR with NNO from group 3, and SVR with NNO and Kriging from group 4.  Figure \ref{fig:final} displays spatial predictions for PM$_{2.5}$ concentrations on June 15, 2019 at 2:00 pm. In particular, panel (d) in Figure \ref{fig:final}, associated with the best performing method, shows a pronounced hotspot in Southern California. The bright yellow patch near ($ 34^{\circ}\mathrm{N},117^{\circ}\mathrm{W}$) corresponds to the Riverside–San Bernardino (Inland Empire) region at the eastern edge of the Los Angeles Basin. Elevated PM$_{2.5}$ in this corridor is consistent with prior literature attributing high levels to basin-confining topography, goods-movement and on-road emissions, episodic inversions/stagnation, and seasonal meteorology (including Santa Ana wind regimes) that modulate temperature, humidity, and wind speed \citep{stowell2020estimating}.
\begin{table}[]
\begin{tabular}{ccccccc}
\hline
Method                 & RMSE & SMAPE & MAD & Cor & 95\% Cov & Time \\ \hline
Group 1: ``Geostatistical" & \multicolumn{5}{c}{}                      \\
UK                    & \textbf{0.3730}  & \textbf{7.802\%}  &  \textbf{0.2630}&  \textbf{0.8701} &  \textbf{93\%} & 3600              \\
NNGP                   & 0.4076 & 9.117\% &0.3174 & 0.7925 &89\% & 5             \\
FRK                    & 0.4227& 9.477\% & 0.3537& 0.7758& 85\% & 420              \\
Group 2: ``Non-Geo"          & \multicolumn{5}{c}{}                      \\
Reg                     & 0.4239&  9.098\% &  0.3102   & 0.4416     & 40\% & 0.1025              \\
RF                     & 0.4927&  10.278\% &  0.3140   & 0.5186     & 67\%  &3.5726             \\

SVR                    &\textbf{0.4186} &\textbf{8.994\%}&\textbf{0.3104} &\textbf{0.6333} &-  &1.5142             \\
ENN                    &0.4725  & 11.825\% & 0.3464 & 0.5412&-&5.6514               \\
Group 3:``Non-Geo+NNO"        & \multicolumn{5}{c}{}                      \\
Reg+NNO                  &0.2608 &4.625\% &  0.1567 & 0.7726 & 90\%  &0.5078                 \\
RF+NNO                  &0.1674 &2.848\% &  0.0964 & 0.9133 & 95\% & 6.0145                  \\
SVR+NNO                &\textbf{0.1552}& \textbf{2.740\%} &\textbf{0.0920}& \textbf{0.9249}& - &     5.1140        \\
ENN+NNO                 & 0.2983& 6.208\%&0.1785 &0.7062&- &6.7755              \\
Group 4:``Non-Geo+NNO+Krig"   & \multicolumn{5}{c}{}                      \\
Reg+NNO+Krig            & 0.0826& 1.390\%& 0.04697& 0.9781& 97\%   &6.0597              \\
RF+NNO+Krig            & 0.0989& 1.458\%& 0.0493& 0.9694& 96\% &12.6580                \\
SVR+NNO+Krig           &\textbf{0.0819} & \textbf{1.330\%} &\textbf{0.0449}& \textbf{0.9792} & - &10.9344               \\
ENN+NNO+Krig            &0.0994       & 2.149\%
      &0.0552  & 0.9690    & -&13
      \end{tabular}
\caption{ A comparison among all models in the four groups with respect to root mean square error (RMSE), symmetric mean absolute percentage error (SMAPE), mean absolute deviation (MAD), the correlation between the observed and predicted values (Cor), 95\% empirical coverage (95\% Cov), and execution time in minutes (Time).}
\label{table:metrics}
\end{table}
\begin{figure}[H]
 \centering
\includegraphics[scale=0.75]{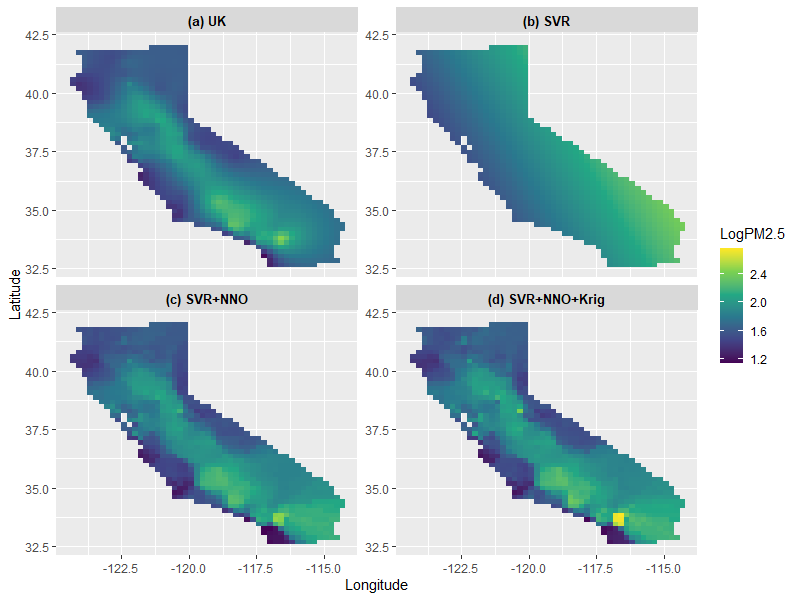}
\caption{\label{fig:final}Predicted log PM$_{2.5}$ concentration for June 15, 2019 at 2:00 pm using (a) Universal Kriging (UK), (b) Support Vector Regression (SVR) with longitude and latitude as predictors, (c) SVR with 10 NNO as predictors and (d) SVR with 10 NNO and Kriging as predictors.}
\end{figure}
\textcolor{black}{\subsection{Sensitivity Analysis: Validation Strategy}}

\textcolor{black}{To assess the sensitivity of model comparisons to the choice of validation strategy, we conducted an additional spatially blocked cross-validation analysis for the hybrid Non-Geo+NNO+Krig models. In contrast to random cross-validation, which allows nearby sensors to appear in both training and test sets, spatial cross-validation withholds entire spatial blocks during model fitting and uses them exclusively for evaluation, thereby reducing spatial dependence between folds.}

\textcolor{black}{Table~\ref{tab:cv-comparison} compares predictive performance under random and spatial cross-validation. As expected, spatial cross-validation results in higher prediction errors and lower correlations across all models, reflecting the increased difficulty of predicting into spatially independent regions rather than interpolating among nearby sensors. Importantly, however, the relative performance ordering of the models remains consistent across validation strategies, with hybrid approaches combining non-spatial predictors, nearest-neighbor operators, and kriging continuing to outperform alternative methods.}

\textcolor{black}{These results indicate that while random cross-validation provides optimistic estimates of absolute performance, the qualitative conclusions regarding comparative model performance are robust to the choice of validation strategy.}

\begin{table}[ht]
\centering
\resizebox{\textwidth}{!}{%
\begin{tabular}{lcccccccccc}
\hline
 & \multicolumn{5}{c}{\textbf{Random CV}} & \multicolumn{5}{c}{\textbf{Spatial CV}} \\
\cline{2-6} \cline{7-11}
\textbf{Method} 
& \textbf{RMSE} & \textbf{SMAPE} & \textbf{MAD} & \textbf{Cor} & \textbf{95\% Cov}
& \textbf{RMSE} & \textbf{SMAPE} & \textbf{MAD} & \textbf{Cor} & \textbf{95\% Cov} \\
\hline
NNGP                   
& 0.4076 & 9.117\% & 0.3174 & 0.7925 & 89\%
& 0.53 & 11.0\% & 0.380 & 0.69 & 92\% \\
Reg+NNO+Krig 
& 0.0826 & 1.390\% & 0.04697 & 0.9781 & 97\%
& 0.18 & 3.5\% & 0.120 & 0.92 & 85\% \\
RF+NNO+Krig  
& 0.0989 & 1.458\% & 0.0493 & 0.9694 & 96\%
& 0.14 & 2.4\% & 0.075 & 0.90 & 94\% \\
SVR+NNO+Krig 
& \textbf{0.0819} & \textbf{1.330\%} & \textbf{0.0449} & \textbf{0.9792} & --
& \textbf{0.12} & \textbf{2.1\%} & \textbf{0.065} & \textbf{0.92} & -- \\
ENN+NNO+Krig 
& 0.0994 & 2.149\% & 0.0552 & 0.9690 & --
& 0.22 & 4.5\% & 0.150 & 0.89 & -- \\
\hline
\end{tabular}
}
\caption{Comparison of hybrid Non-Geo+NNO+Krig models under random cross-validation (Random CV) and spatial cross-validation (Spatial CV).}
\label{tab:cv-comparison}
\end{table}

\section{Conclusions and Discussion}
This study provides a comprehensive assessment of prevalent geostatistical and machine learning methods for estimating PM$_{2.5}$ concentrations across spatial and temporal scales, utilizing an extensive network of PurpleAir sensors to enhance prediction accuracy for air quality in California. 

While geostatistical methods effectively capture spatiotemporal dependencies, they are constrained by high dimensionality and computational demands. For instance, Universal Kriging  required up to 3600 minutes, compared to 420 minutes for Fixed Rank Kriging and only 5 minutes for the Nearest Neighbor Gaussian Process. For the non-geostatistical machine learning algorithms, the maximum execution time was about 6 minutes recorded for ensemble neural network. Machine learning algorithms offer rapid implementation and robustness to non-linear relationships, but they generally overlook spatiotemporal dependencies and uncertainty. 

Our findings highlight the synergistic potential of integrating geostatistical approaches with machine learning to enhance prediction accuracy without relying on assumptions of stationarity, linearity, or normality. Particularly, incorporating predictors from the fastest geostatistical method in our comparison, Nearest Neighbor Gaussian Process, into machine learning models significantly improved prediction quality, reduced RMSE, SMAPE, and MAD, and increased the correlation between observed and predicted PM$_{2.5}$ values. \textcolor{black}{Among the hybrid models, SVR+NNO+Krig achieved the lowest point-prediction error across evaluation metrics. However, models such as Reg+NNO+Krig and RF+NNO+Krig provided comparable predictive accuracy while also delivering calibrated uncertainty estimates, which may be preferable in operational or decision-support contexts.} Overall, our comparison addresses the often-neglected combined spatiotemporal dynamics, which are critical for understanding and managing air quality. More importantly, our findings reveal significant advantages when integrating geostatistical methods with machine learning algorithms. 

Despite these strengths, several limitations should be acknowledged. First, we used the Barkjohn et al.\ (2020) calibration formula for correcting PurpleAir data, which, although widely adopted, may not fully capture spatial or humidity-driven variation in sensor bias. More recent dynamic or site-specific calibration models could further improve data quality, particularly during wildfire events or high-humidity periods. Second, while our primary evaluation focused on short-range interpolation accuracy using random spatial cross-validation within a dense sensor network, \textcolor{black}{we additionally conducted spatially blocked cross-validation as a sensitivity analysis. As expected, spatial blocking produced higher error metrics and lower correlations, reflecting the increased difficulty of predicting into spatially independent regions rather than interpolating among nearby sensors. However, the relative performance ordering of the models was preserved, indicating that the comparative advantages of the hybrid framework are robust to the choice of validation strategy. Nevertheless, absolute performance under spatial generalization remains more conservative than under interpolation settings, and applications targeting geographically distinct regions may require validation schemes tailored to broader extrapolation contexts.} Third, while deterministic machine learning models such as Support Vector Regression and Ensemble Neural Network demonstrated excellent point predictive accuracy, they lack intrinsic uncertainty quantification. Future extensions could explore probabilistic machine learning formulations, such as Bayesian support vector regression or Monte Carlo dropout neural networks, to provide calibrated predictive uncertainty alongside point predictions and improve interpretability.

Incorporating additional environmental and anthropogenic covariates, such as meteorological variables, land-use characteristics, and emissions-related indicators, could further enhance predictive performance and improve the identification of pollution hotspots \citep{bao2016association, perez2010variability, hoek2008review}. Relative to recent advances in spatiotemporal deep learning and data fusion approaches that integrate satellite retrievals, chemical transport models, and sensor data, our hybrid geostatistical machine learning framework offers a complementary perspective. While large-scale data fusion methods often prioritize broad spatial generalization, our framework emphasizes superior local interpolation performance within dense sensor networks and, for models providing predictive variance (e.g., geostatistical and RF-based approaches), interpretable spatial uncertainty quantification.

Looking ahead, future studies could explore the integration of more diverse data sources, including real-time traffic and industrial emissions data, to further contextualize the spatiotemporal dynamics of air pollution. Additionally, advancing algorithmic approaches to more effectively process large-scale environmental data could provide deeper insights into the complex interactions affecting air quality. Such advancements will not only improve predictive accuracy but also enhance our understanding of the environmental and anthropogenic factors driving PM$_{2.5}$ variability, supporting more targeted and effective air quality management strategies. When combined with probabilistic modeling frameworks that provide uncertainty quantification, the integration of richer environmental covariates and large-scale data streams would further enhance model interpretability and decision relevance for air-quality management.

\footnotesize

\section*{Acknowledgments}

The authors would like to thank Tamer Oraby, Yawen Guan for helpful discussions. We also appreciate the effort of the Editor, Associate Editor and Referees to improve quality of the manuscript.

\appendix
\section{Appendix}
\begin{figure}[H]
\begin{subfigure}{.5\textwidth}
\caption*{(a) \textbf{UK}}
\includegraphics[width=\linewidth]{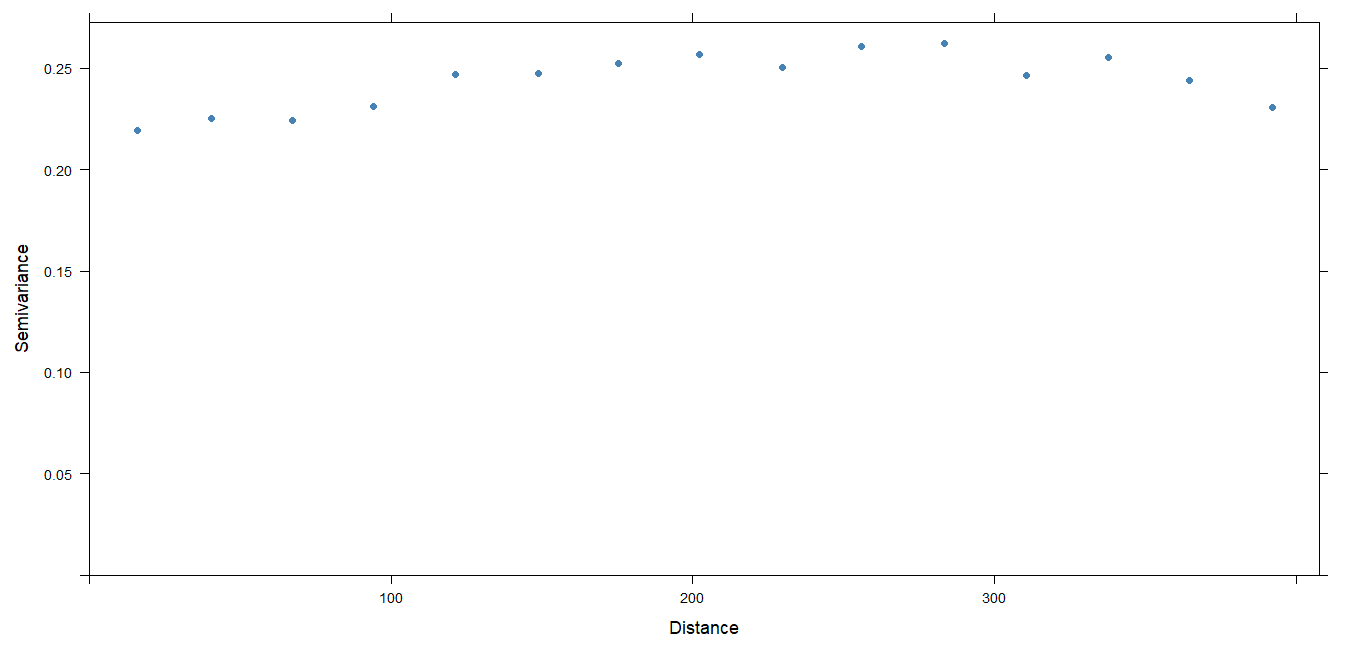}
\end{subfigure}
\begin{subfigure}{.5\textwidth}
\caption*{(b) \textbf{NNGP}}
\includegraphics[width=\linewidth]{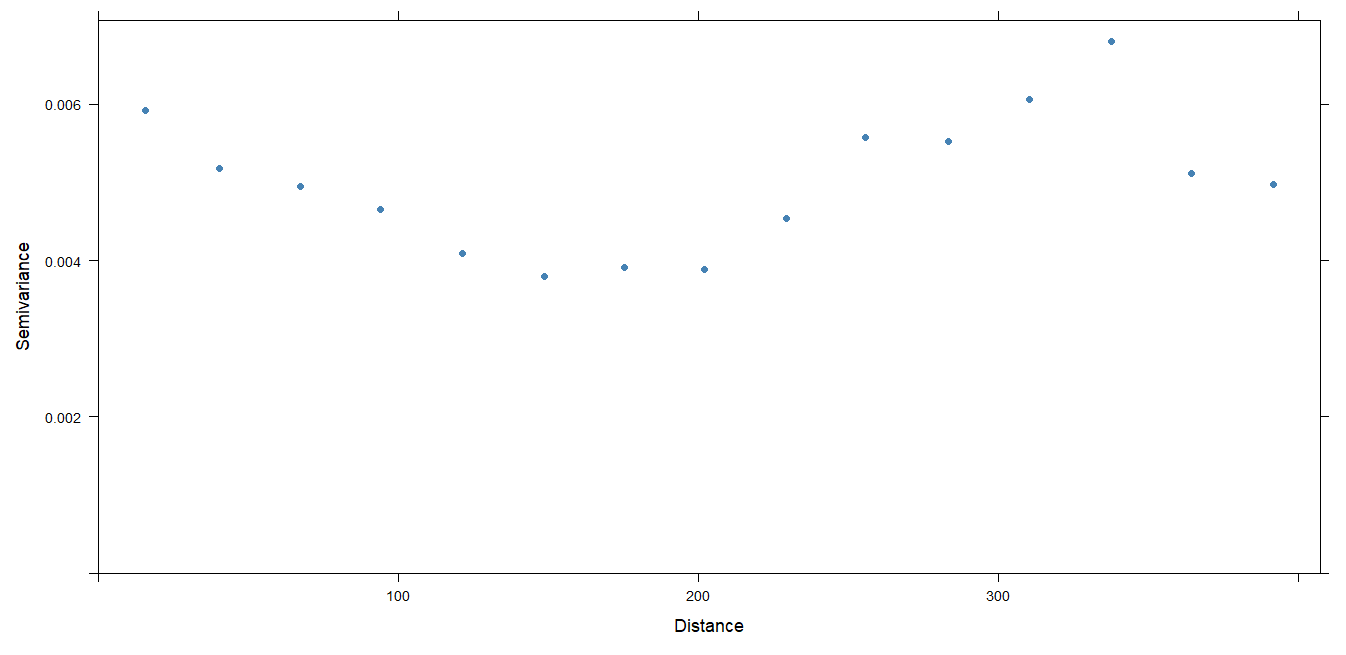}
\end{subfigure}
\begin{subfigure}{.5\textwidth}
\caption*{(c) \textbf{FRK}}
\includegraphics[width=\linewidth]{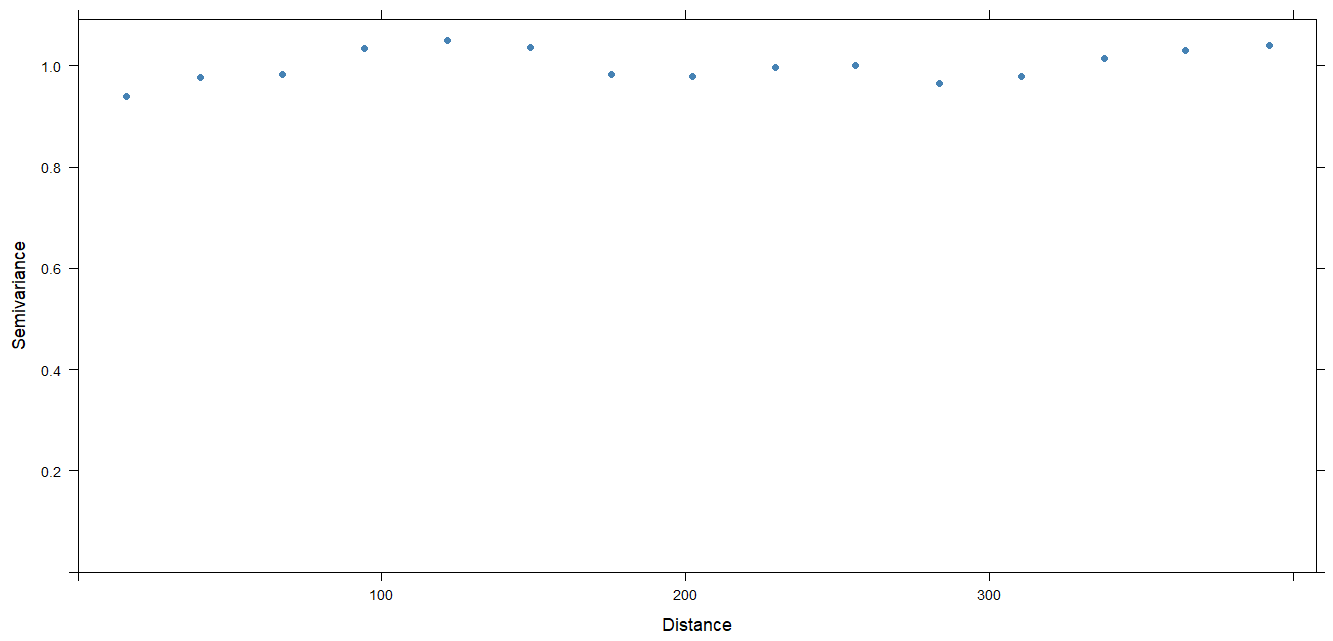}
\end{subfigure}
\begin{subfigure}{.5\textwidth}
\caption*{(d) \textbf{Reg+NNO+Krig}}
\includegraphics[width=\linewidth]{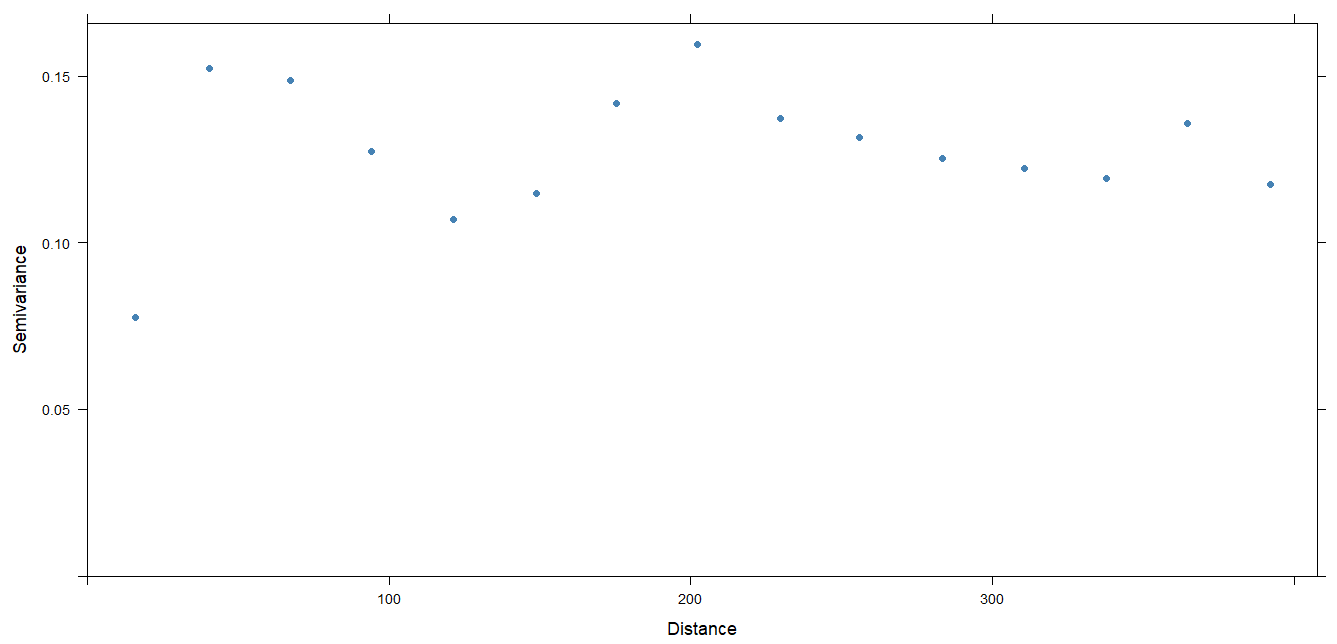}
\end{subfigure}
\begin{subfigure}{.5\textwidth}
\caption*{(e) \textbf{RF+NNO+Krig}}
\includegraphics[width=\linewidth]{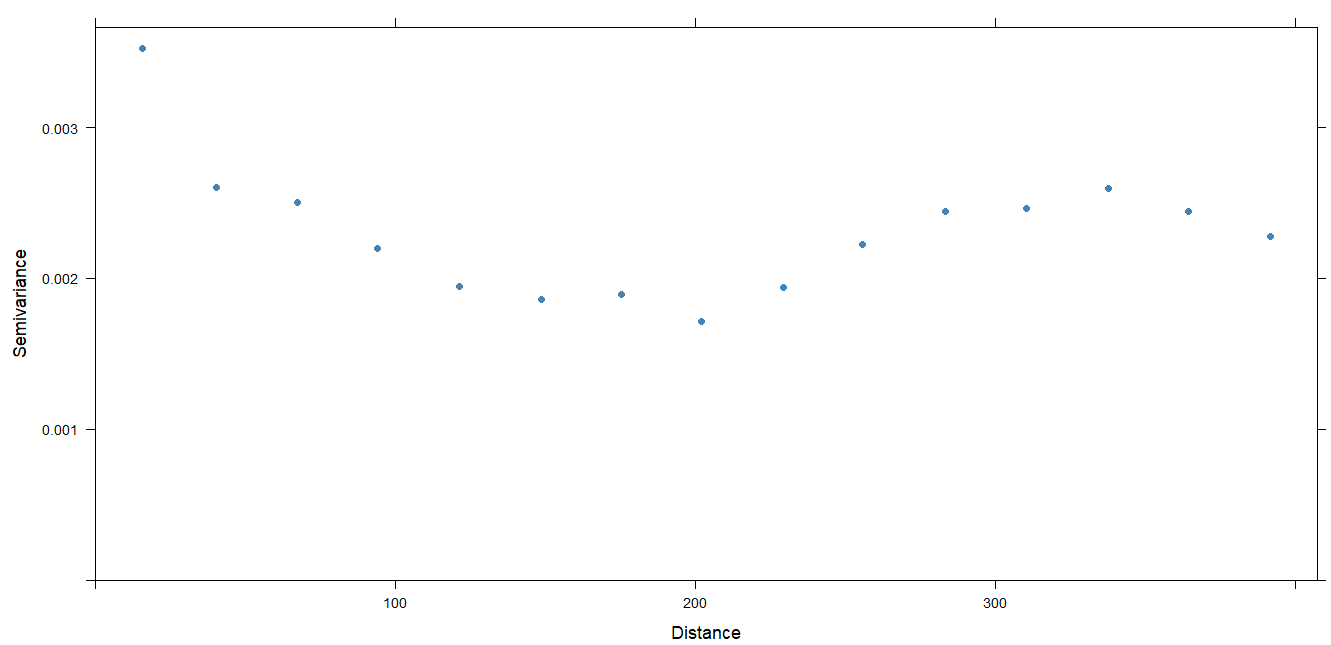}
\end{subfigure}
\begin{subfigure}{.5\textwidth}
\caption*{(f) \textbf{SVR+NNO+Krig}}
\includegraphics[width=\linewidth]{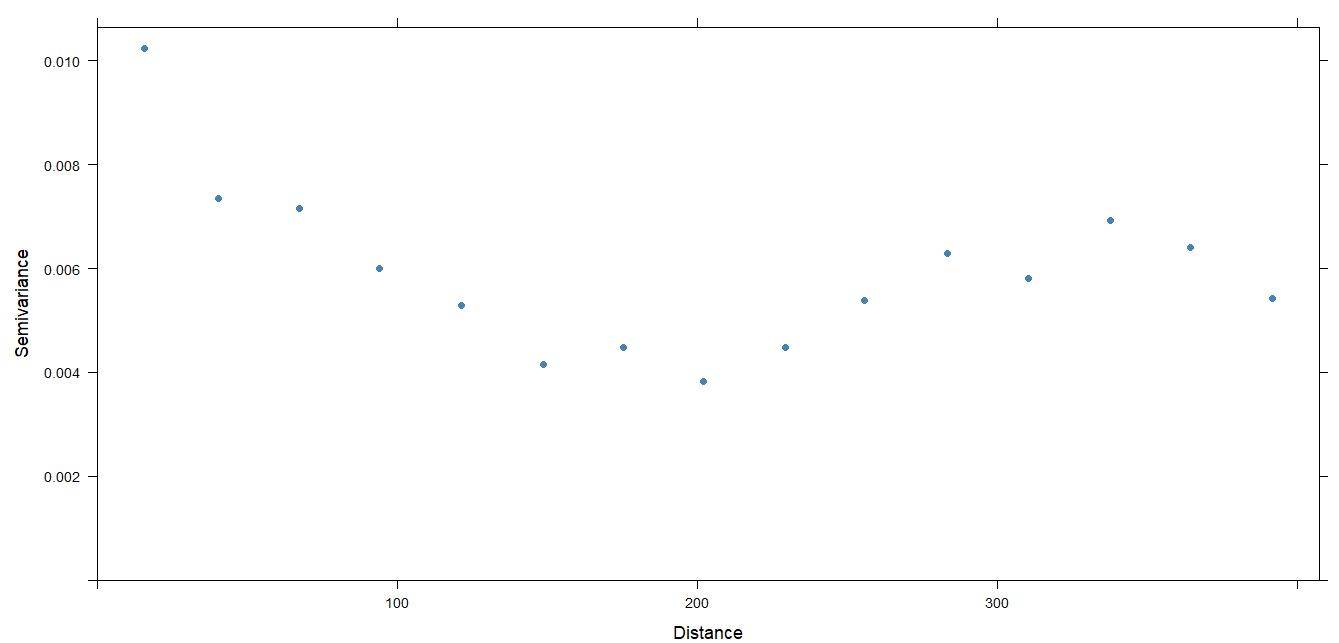}
\end{subfigure}
\caption{Residual semivariogram plots for (a) UK, (b) NNGP, (c) FRK, (d) Reg+NNO+Krig, (e) RF+NNO+Krig, and (f) SVR+NNO+Krig.}
\label{fig:r2-ce}
\end{figure}
\bibliography{references3.bib}

@article{gong2021multivariate,
  title   = {Multivariate spatial prediction of air pollutant concentrations with {INLA}},
  author  = {Gong, Wenlong and Reich, Brian J. and Chang, Howard H.},
  journal = {Environmental Research Communications},
  volume  = {3},
  number  = {10},
  pages   = {101002},
  year    = {2021}
}

@article{Chen2023,
  title   = {Estimation of fine-resolution {PM2.5} concentrations using the {INLA-SPDE} method},
  author  = {Chen, Jingna and Miao, Changhong and Yang, Dongyang and Liu, Yong and Zhang, Hang and Dong, Guanpeng},
  journal = {Atmospheric Pollution Research},
  volume  = {14},
  number  = {7},
  pages   = {101781},
  year    = {2023},
  doi     = {10.1016/j.apr.2023.101781}
}

@techreport{world2018noncommunicable,
  author      = {{World Health Organization}},
  title       = {Noncommunicable Diseases Country Profiles 2018},
  institution = {World Health Organization},
  year        = {2018},
  date        = {2018-09-24},
  type        = {Technical document},
  number      = {ISBN 978-92-4-151462-0},
  pages       = {223},
  url         = {https://www.who.int/publications/i/item/ncd-country-profiles-2018}
}

@misc{GpGp,
  title        = {Fast {G}aussian Process Computation Using Vecchia's Approximation: R Package \texttt{GpGp} (Version 0.4.0)},
  author       = {Guinness, Joseph and Katzfuss, Matthias and Fahmy, Youssef},
  year         = {2021},
  howpublished = {R package}
}

@article{hajat2015socioeconomic,
  title   = {Socioeconomic disparities and air pollution exposure: a global review},
  author  = {Hajat, Anjum and Hsia, Charlene and O’Neill, Marie S.},
  journal = {Current Environmental Health Reports},
  volume  = {2},
  pages   = {440--450},
  year    = {2015}
}

@article{morawska2018applications,
  title   = {Applications of low-cost sensing technologies for air quality monitoring and exposure assessment: How far have they gone?},
  author  = {Morawska, Lidia and Thai, Phong K. and Liu, Xiaoting and Asumadu-Sakyi, Akwasi and Ayoko, Godwin and Bartonova, Alena and Bedini, Andrea and Chai, Fahe and Christensen, Bryce and Dunbabin, Matthew and others},
  journal = {Environment International},
  volume  = {116},
  pages   = {286--299},
  year    = {2018}
}

@article{bi2020incorporating,
  title   = {Incorporating low-cost sensor measurements into high-resolution {PM2.5} modeling at a large spatial scale},
  author  = {Bi, Jianzhao and Wildani, Avani and Chang, Howard H. and Liu, Yang},
  journal = {Environmental Science \& Technology},
  volume  = {54},
  number  = {4},
  pages   = {2152--2162},
  year    = {2020}
}

@article{delp2020wildfire,
  title   = {Wildfire smoke adjustment factors for low-cost and professional {PM2.5} monitors with optical sensors},
  author  = {Delp, William W. and Singer, Brett C.},
  journal = {Sensors},
  volume  = {20},
  number  = {13},
  pages   = {3683},
  year    = {2020}
}

@article{alexeeff2015consequences,
  title   = {Consequences of {k}riging and land use regression for {PM2.5} predictions in epidemiologic analyses: insights into spatial variability using high-resolution satellite data},
  author  = {Alexeeff, Stacey E. and Schwartz, Joel and Kloog, Itai and Chudnovsky, Alexandra and Koutrakis, Petros and Coull, Brent A.},
  journal = {Journal of Exposure Science \& Environmental Epidemiology},
  volume  = {25},
  number  = {2},
  pages   = {138--144},
  year    = {2015}
}

@inproceedings{wang2019nearest,
  title     = {Nearest-neighbor neural networks for geostatistics},
  author    = {Wang, Haoyu and Guan, Yawen and Reich, Brian},
  booktitle = {2019 International Conference on Data Mining Workshops (ICDMW)},
  pages     = {196--205},
  year      = {2019},
  organization = {IEEE}
}

@article{datta2016hierarchical,
  title   = {Hierarchical nearest-neighbor {G}aussian process models for large geostatistical datasets},
  author  = {Datta, Abhirup and Banerjee, Sudipto and Finley, Andrew O. and Gelfand, Alan E.},
  journal = {Journal of the American Statistical Association},
  volume  = {111},
  number  = {514},
  pages   = {800--812},
  year    = {2016}
}

@article{hu2013estimating,
  title   = {Estimating ground-level {PM2.5} concentrations in the southeastern US using geographically weighted regression},
  author  = {Hu, Xuefei and Waller, Lance A. and Al-Hamdan, Mohammad Z. and Crosson, William L. and Estes Jr, Maurice G. and Estes, Sue M. and Quattrochi, Dale A. and Sarnat, Jeremy A. and Liu, Yang},
  journal = {Environmental Research},
  volume  = {121},
  pages   = {1--10},
  year    = {2013}
}

@article{gupta2009particulate,
  title   = {Particulate matter air quality assessment using integrated surface, satellite, and meteorological products: 2. {A} neural network approach},
  author  = {Gupta, Pawan and Christopher, Sundar A.},
  journal = {Journal of Geophysical Research: Atmospheres},
  volume  = {114},
  number  = {D20},
  year    = {2009}
}

@article{hu2017estimating,
  title   = {Estimating {PM2.5} concentrations in the conterminous {U}nited {S}tates using the random forest approach},
  author  = {Hu, Xuefei and Belle, Jessica H. and Meng, Xia and Wildani, Avani and Waller, Lance A. and Strickland, Matthew J. and Liu, Yang},
  journal = {Environmental Science \& Technology},
  volume  = {51},
  number  = {12},
  pages   = {6936--6944},
  year    = {2017}
}

@article{mogollon2021support,
  title   = {A support vector machine model to forecast ground-level {PM2.5} in a highly populated city with a complex terrain},
  author  = {Mogoll{\'o}n-Sotelo, Caroline and Casallas, Alejandro and Vidal, Sergio and Celis, Nathalia and Ferro, Camilo and Belalcazar, Luis},
  journal = {Air Quality, Atmosphere \& Health},
  volume  = {14},
  pages   = {399--409},
  year    = {2021}
}

@article{requia2019evaluation,
  title   = {Evaluation of predictive capabilities of ordinary geostatistical interpolation, hybrid interpolation, and machine learning methods for estimating {PM2.5} constituents over space},
  author  = {Requia, Weeberb J. and Coull, Brent A. and Koutrakis, Petros},
  journal = {Environmental Research},
  volume  = {175},
  pages   = {421--433},
  year    = {2019}
}

@article{berrocal2020comparison,
  title   = {A comparison of statistical and machine learning methods for creating national daily maps of ambient {PM2.5} concentration},
  author  = {Berrocal, Veronica J. and Guan, Yawen and Muyskens, Amanda and Wang, Haoyu and Reich, Brian J. and Mulholland, James A. and Chang, Howard H.},
  journal = {Atmospheric Environment},
  volume  = {222},
  pages   = {117130},
  year    = {2020}
}

@article{maliva2016geostatistical,
  title   = {Geostatistical Methods and Applications},
  author  = {Maliva, Robert G.},
  journal = {Aquifer Characterization Techniques: Schlumberger Methods in Water Resources Evaluation Series No. 4},
  pages   = {595--617},
  year    = {2016},
  publisher = {Springer}
}

@article{heaton2019case,
  title   = {A case study competition among methods for analyzing large spatial data},
  author  = {Heaton, Matthew J. and Datta, Abhirup and Finley, Andrew O. and Furrer, Reinhard and Guinness, Joseph and Guhaniyogi, Rajarshi and Gerber, Florian and Gramacy, Robert B. and Hammerling, Dorit and Katzfuss, Matthias and others},
  journal = {Journal of Agricultural, Biological and Environmental Statistics},
  volume  = {24},
  pages   = {398--425},
  year    = {2019}
}

@article{datta2016nonseparable,
  title   = {Nonseparable dynamic nearest neighbor {G}aussian process models for large spatio-temporal data with an application to particulate matter analysis},
  author  = {Datta, Abhirup and Banerjee, Sudipto and Finley, Andrew O. and Hamm, Nicholas A. S. and Schaap, Martijn and Gelfand, Alan E.},
  journal = {The Annals of Applied Statistics},
  volume  = {10},
  number  = {3},
  pages   = {1286},
  year    = {2016}
}

@article{breiman2001statistical,
  title   = {Statistical modeling: The two cultures},
  author  = {Breiman, Leo},
  journal = {Statistical Science},
  volume  = {16},
  number  = {3},
  pages   = {199--231},
  year    = {2001}
}

@article{di2016assessing,
  title   = {Assessing {PM2.5} exposures with high spatiotemporal resolution across the continental {U}nited {S}tates},
  author  = {Di, Qian and Kloog, Itai and Koutrakis, Petros and Lyapustin, Alexei and Wang, Yujie and Schwartz, Joel},
  journal = {Environmental Science \& Technology},
  volume  = {50},
  number  = {9},
  pages   = {4712--4721},
  year    = {2016}
}

@article{reid2015spatiotemporal,
  title   = {Spatiotemporal prediction of fine particulate matter during the 2008 northern California wildfires using machine learning},
  author  = {Reid, Colleen E. and Jerrett, Michael and Petersen, Maya L. and Pfister, Gabriele G. and Morefield, Philip E. and Tager, Ira B. and Raffuse, Sean M. and Balmes, John R.},
  journal = {Environmental Science \& Technology},
  volume  = {49},
  number  = {6},
  pages   = {3887--3896},
  year    = {2015}
}

@inproceedings{athmaja2017survey,
  title     = {A survey of machine learning algorithms for big data analytics},
  author    = {Athmaja, S. and Hanumanthappa, M. and Kavitha, Vasantha},
  booktitle = {2017 International Conference on Innovations in Information, Embedded and Communication Systems (ICIIECS)},
  pages     = {1--4},
  year      = {2017},
  organization = {IEEE}
}

@book{vapnik1999nature,
  title     = {The Nature of Statistical Learning Theory},
  author    = {Vapnik, Vladimir},
  year      = {1999},
  publisher = {Springer Science \& Business Media}
}

@misc{vapnik1974theory,
  title     = {Theory of Pattern Recognition},
  author    = {Vapnik, Vladimir and Chervonenkis, Alexey},
  year      = {1974},
  publisher = {Nauka, Moscow}
}

@inproceedings{Drucker1997,
  author    = {Drucker, Harris and Surges, Chris J. C. and Kaufman, Linda and Smola, Alex and Vapnik, Vladimir},
  booktitle = {Advances in Neural Information Processing Systems},
  isbn      = {0262100657},
  issn      = {10495258},
  title     = {Support Vector Regression Machines},
  year      = {1997}
}

@article{stowell2020estimating,
  title   = {Estimating {PM2.5} in Southern California using satellite data: factors that affect model performance},
  author  = {Stowell, Jennifer D. and Bi, Jianzhao and Al-Hamdan, Mohammad Z. and Lee, Hyung Joo and Lee, Sang-Mi and Freedman, Frank and Kinney, Patrick L. and Liu, Yang},
  journal = {Environmental Research Letters},
  volume  = {15},
  number  = {9},
  pages   = {094004},
  year    = {2020}
}

@article{perez2010variability,
  title   = {Variability of particle number, black carbon, and {PM10}, {PM2.5}, and {PM1} levels and speciation: influence of road traffic emissions on urban air quality},
  author  = {P{\'e}rez, Noem{\'\i} and Pey, Jorge and Cusack, Michael and Reche, Cristina and Querol, Xavier and Alastuey, Andr{\'e}s and Viana, Mar},
  journal = {Aerosol Science and Technology},
  volume  = {44},
  number  = {7},
  pages   = {487--499},
  year    = {2010}
}

@article{bao2016association,
  title   = {Association of $PM_{2.5}$ pollution with the pattern of human activity: A case study of a developed city in eastern {C}hina},
  author  = {Bao, Chengzhen and Chai, Pengfei and Lin, Hongbo and Zhang, Zhenyu and Ye, Zhenhua and Gu, Mengjia and Lu, Huaichu and Shen, Peng and Jin, Mingjuan and Wang, Jianbing and others},
  journal = {Journal of the Air \& Waste Management Association},
  volume  = {66},
  number  = {12},
  pages   = {1202--1213},
  year    = {2016}
}

@article{hoek2008review,
  title   = {A review of land-use regression models to assess spatial variation of outdoor air pollution},
  author  = {Hoek, Gerard and Beelen, Rob and De Hoogh, Kees and Vienneau, Danielle and Gulliver, John and Fischer, Paul and Briggs, David},
  journal = {Atmospheric Environment},
  volume  = {42},
  number  = {33},
  pages   = {7561--7578},
  year    = {2008}
}

@article{liaw2002classification,
  title   = {Classification and regression by randomForest},
  author  = {Liaw, Andy and Wiener, Matthew and others},
  journal = {R News},
  volume  = {2},
  number  = {3},
  pages   = {18--22},
  year    = {2002}
}

@article{zammit2021frk,
  title   = {FRK: An R package for spatial and spatio-temporal prediction with large datasets},
  author  = {Zammit-Mangion, Andrew and Cressie, Noel},
  journal = {Journal of Statistical Software},
  volume  = {98},
  pages   = {1--48},
  year    = {2021}
}

@article{dimitriadou2008misc,
  title   = {Misc Functions of the Department of Statistics (e1071), TU Wien},
  author  = {Dimitriadou, Evgenia and Hornik, Kurt and Leisch, Friedrich and Meyer, David and Weingessel, Andreas},
  journal = {R Package},
  volume  = {1},
  pages   = {5--24},
  year    = {2008}
}

@article{hansen1990neural,
  title   = {Neural network ensembles},
  author  = {Hansen, Lars Kai and Salamon, Peter},
  journal = {IEEE Transactions on Pattern Analysis and Machine Intelligence},
  volume  = {12},
  number  = {10},
  pages   = {993--1001},
  year    = {1990}
}

@article{graler2016spatio,
  title   = {Spatio-temporal interpolation using gstat},
  author  = {Gr{\"a}ler, Benedikt and Pebesma, Edzer J. and Heuvelink, Gerard B. M.},
  journal = {The R Journal},
  volume  = {8},
  number  = {1},
  pages   = {204},
  year    = {2016}
}

@online{purpleair,
  author  = {{PurpleAir}},
  title   = {{P}urple{A}ir - Real Time Air Quality Monitoring},
  year    = {2021},
  url     = {https://www.purpleair.com/},
  urldate = {2023-03-28}
}

@misc{API,
  title        = {Airsensor},
  howpublished = {\url{https://api.purpleair.com/}},
  note         = {Accessed: 04 06, 2023}
}

@manual{R,
  title        = {R: A Language and Environment for Statistical Computing},
  author       = {{R Core Team}},
  organization = {R Foundation for Statistical Computing},
  address      = {Vienna, Austria},
  year         = {2021},
  url          = {https://www.R-project.org/}
}

@online{chollet2015keras,
  title     = {Keras},
  author    = {Chollet, Francois and others},
  year      = {2015},
  publisher = {GitHub},
  url       = {https://github.com/fchollet/keras}
}

@article{pedregosa2011scikit,
  title   = {Scikit-learn: Machine learning in Python},
  author  = {Pedregosa, Fabian and Varoquaux, Ga{\"e}l and Gramfort, Alexandre and Michel, Vincent and Thirion, Bertrand and Grisel, Olivier and Blondel, Mathieu and Prettenhofer, Peter and Weiss, Ron and Dubourg, Vincent and others},
  journal = {Journal of Machine Learning Research},
  volume  = {12},
  number  = {Oct},
  pages   = {2825--2830},
  year    = {2011}
}

@book{van1995python,
  title     = {Python Reference Manual},
  author    = {Van Rossum, Guido and Drake Jr, Fred L.},
  year      = {1995},
  publisher = {Centrum voor Wiskunde en Informatica Amsterdam}
}

@article{barkjohn2020development,
  title   = {Development and application of a {U}nited {S}tates-wide correction for {PM2.5} data collected with the {P}urple{A}ir sensor},
  author  = {Barkjohn, Karoline K. and Gantt, Brett and Clements, Andrea L.},
  journal = {Atmospheric Measurement Techniques},
  year    = {2020}
}

@book{wikle2019spatio,
  title     = {Spatio-Temporal Statistics with {R}},
  author    = {Wikle, Christopher K. and Zammit-Mangion, Andrew and Cressie, Noel},
  year      = {2019},
  publisher = {Chapman and Hall/CRC}
}

@article{cressie2008fixed,
  title   = {Fixed rank kriging for very large spatial data sets},
  author  = {Cressie, Noel and Johannesson, Gardar},
  journal = {Journal of the Royal Statistical Society: Series B (Statistical Methodology)},
  volume  = {70},
  number  = {1},
  pages   = {209--226},
  year    = {2008}
}

@article{vecchia1988estimation,
  title   = {Estimation and model identification for continuous spatial processes},
  author  = {Vecchia, Aldo V.},
  journal = {Journal of the Royal Statistical Society: Series B (Methodological)},
  volume  = {50},
  number  = {2},
  pages   = {297--312},
  year    = {1988}
}

@book{hastie2009elements,
  title     = {The Elements of Statistical Learning},
  author    = {Hastie, Trevor and Tibshirani, Robert and Friedman, Jerome},
  edition   = {2},
  year      = {2009},
  publisher = {Springer}
}

\end{justify}

\end{document}